\documentstyle[epsfig]{mn}

\newif\ifAMStwofonts

\ifoldfss
  \ifCUPmtlplainloaded \else
    \NewTextAlphabet{textbfit} {cmbxti10} {}
    \NewTextAlphabet{textbfss} {cmssbx10} {}
    \NewMathAlphabet{mathbfit} {cmbxti10} {} 
    \NewMathAlphabet{mathbfss} {cmssbx10} {} 
  \fi
  \ifAMStwofonts
    \ifCUPmtlplainloaded \else
      \NewSymbolFont{upmath} {eurm10}
      \NewSymbolFont{AMSa} {msam10}
      \NewMathSymbol{\upi}     {0}{upmath}{19}
      \NewMathSymbol{\umu}     {0}{upmath}{16}
      \NewMathSymbol{\upartial}{0}{upmath}{40}
      \NewMathSymbol{\leqslant}{3}{AMSa}{36}
      \NewMathSymbol{\geqslant}{3}{AMSa}{3E}

    \fi
  \fi
\fi 

\ifnfssone
  \newmathalphabet{\mathit}
  \addtoversion{normal}{\mathit}{cmr}{m}{it}
  \addtoversion{bold}{\mathit}{cmr}{bx}{it}
  \newmathalphabet{\mathbfit} 
  \addtoversion{normal}{\mathbfit}{cmr}{bx}{it}
  \addtoversion{bold}{\mathbfit}{cmr}{bx}{it}
  \newmathalphabet{\mathbfss} 
  \addtoversion{normal}{\mathbfss}{cmss}{bx}{n}
  \addtoversion{bold}{\mathbfss}{cmss}{bx}{n}
  \ifAMStwofonts
    \ifCUPmtlplainloaded \else
      \UseAMStwoboldmath
      \makeatletter
      \new@mathgroup\upmath@group
      \define@mathgroup\mv@normal\upmath@group{eur}{m}{n}
      \define@mathgroup\mv@bold\upmath@group{eur}{b}{n}
      \edef\UPM{\hexnumber\upmath@group}
      \new@mathgroup\amsa@group
      \define@mathgroup\mv@normal\amsa@group{msa}{m}{n}
      \define@mathgroup\mv@bold\amsa@group{msa}{m}{n}
      \edef\AMSa{\hexnumber\amsa@group}
      \makeatother
      \mathchardef\upi="0\UPM19
      \mathchardef\umu="0\UPM16
      \mathchardef\upartial="0\UPM40
      \mathchardef\leqslant="3\AMSa36
      \mathchardef\geqslant="3\AMSa3E
    \fi
  \fi
\fi 

\ifnfsstwo
  \DeclareMathAlphabet{\mathbfit}{OT1}{cmr}{bx}{it}
  \SetMathAlphabet\mathbfit{bold}{OT1}{cmr}{bx}{it}
  \DeclareMathAlphabet{\mathbfss}{OT1}{cmss}{bx}{n}
  \SetMathAlphabet\mathbfss{bold}{OT1}{cmss}{bx}{n}
  \ifAMStwofonts
    \ifCUPmtlplainloaded \else
      \DeclareSymbolFont{UPM}{U}{eur}{m}{n}
      \SetSymbolFont{UPM}{bold}{U}{eur}{b}{n}
      \DeclareSymbolFont{AMSa}{U}{msa}{m}{n}
      \DeclareMathSymbol{\upi}{0}{UPM}{"19}
      \DeclareMathSymbol{\umu}{0}{UPM}{"16}
      \DeclareMathSymbol{\upartial}{0}{UPM}{"40}
      \DeclareMathSymbol{\leqslant}{3}{AMSa}{"36}
      \DeclareMathSymbol{\geqslant}{3}{AMSa}{"3E}
    \fi
  \fi
\fi 

\ifCUPmtlplainloaded \else
  \ifAMStwofonts \else 
    \def\upi{\pi}
    \def\umu{\mu}
    \def\upartial{\partial}
  \fi
\fi

\title[First Results from the H{\sc I} Jodrell All Sky Survey]{First Results from the H{\sc I} Jodrell 
 All Sky Survey: Inclination-Dependent Selection Effects in a 21-cm Blind Survey}
\author[R.H. Lang et al.]
       {Robert H. Lang$^1$, Peter J. Boyce,$^2$, Virginia A. Kilborn$^3$, Robert F. Minchin$^1$,  
 \newauthor Michael J. Disney$^1$, Christine A. Jordan$^3$, Marco Grossi$^1$, 
  Diego A. Garcia$^1$, \newauthor Ken C. Freeman$^4$, Steven Phillipps$^2$ and Alan E. Wright$^5$  \\ 
   $^1$Department of Physics and Astronomy, Cardiff University, P.O. Box 913, Cardiff, CF24 3YB \\
 $^2$Astrophysics Group, Department of Physics, University of Bristol, Tyndall Avenue, Bristol, BS8 1TL\\
     $^3$Jodrell Bank Observatory, University of Manchester, Macclesfield, 
 Cheshire, SK11 9DL\\  
$^4$Research School of Astronomy and Astrophysics, Mount Stromlo Observatory, Cotter Road, Weston, ACT 1611, Australia\\
$^5$Australia Telescope National Facility, CSIRO, P.O. Box 76, Epping, NSW 1710, Australia  }

\date{Accepted ???.
      Received ???? ;}

\pagerange{\pageref{firstpage}--\pageref{lastpage}}
\pubyear{1994}

\begin{document}

\maketitle

\label{firstpage}

\begin{abstract}
Details are presented of the H{\sc I} Jodrell All Sky Survey (H{\sc I}JASS). 
H{\sc I}JASS is a blind  neutral hydrogen (H{\sc I}) survey 
of the northern sky ($\delta$$>$22\degr), being conducted 
 using the   multibeam receiver on the Lovell 
 Telescope (FWHM beamwidth 12~arcmin) at Jodrell Bank. 
 H{\sc I}JASS covers the velocity range --3500~km\,s$^{-1}$
 to 10000~km\,s$^{-1}$, with a velocity resolution of 18.1~km\,s$^{-1}$
 and spatial positional accuracy of $\sim$2.5~arcmin. 
 Thus far about 1115~deg$^{2}$ of sky  have been surveyed. 
 The average rms noise  during the early part of the survey was  
  around 16~mJy~beam$^{-1}$. Following the first phase of the Lovell 
 telescope upgrade (in 2001), the rms noise is now 
 around 13~mJy~beam$^{-1}$. We describe 
   the methods of detecting galaxies within the H{\sc I}JASS data and of measuring their 
 H{\sc I} parameters. The properties of the resulting H{\sc I}-selected sample 
 of galaxies are described. 
    Of the 222 sources so far confirmed, 170 (77 per cent) are clearly associated 
 with a previously catalogued galaxy.  A further 23 sources
 (10 per cent) lie close (within 6~arcmin) to a previously 
  catalogued galaxy for which no previous redshift 
 exists. A further 29 sources (13 per cent) do not appear to be associated with
 any previously catalogued galaxy.   The distributions of 
 peak flux, integrated flux, H{\sc I} mass and $cz$ are discussed.  
  We show, using the H{\sc I}JASS data, that H{\sc I} self-absorption is a significant, but often overlooked, effect in galaxies with large inclination angles 
 to the   line of sight. 
Properly accounting for it could increase the derived HI mass density of the local Universe by at 
 least 25 per cent. The effect this will have on the shape 
 of the H{\sc I} Mass Function (H{\sc I}MF) will depend on how self-absorption affects galaxies of different morphological types and 
 H{\sc I} masses.  We  also show that galaxies with small inclinations to the line of sight 
  may also be excluded from HI-selected samples, since many 
 such galaxies will have observed velocity-widths which are too narrow for them to be distinguished from narrow-band radio 
 frequency interference. 
 This  effect will become progressively more serious  for galaxies with smaller intrinsic  
 velocity-widths. If, as we might expect, galaxies with smaller
 intrinsic velocity-widths have smaller H{\sc I} masses, then compensating for this effect could 
 significantly steepen the   
  faint-end slope of the derived H{\sc I}MF.
 
\end{abstract}


\section{Introduction}

A complete and bias-free census of the 
 population of extragalactic objects is essential to any 
 study of the formation and evolution of galaxies or the 
 large-scale structure of the universe. However, our 
 current understanding of galaxy populations 
 has been primarily derived from optical 
 and IR surveys. There is an inevitable bias in such surveys 
 against low luminosity objects (dwarfs), but also 
 against  low surface brightness (LSB) objects
    (see, e.g.  Disney 1976; Disney \& Phillipps 1987; Impey \& Bothun 1997; Disney 1999).

However, 
 it has  become clear that low luminosity and low
surface brightness  galaxies play a key 
role in many cosmological and cosmographical problems.
  For example, dwarf/LSB galaxies  can clearly
play a major role in helping us to understand large-scale structure and
its influence on galaxy formation and evolution. Their
numbers and distribution 
place constraints on the increasingly sophisticated numerical and
semi-analytic models of galaxy formation (e.g., Baugh, Cole \& Frenk 1996; 
Kauffmann et al. 1997), while their morphologies
and stellar contents may reflect the local physics which define
the star formation process in galaxies (e.g., Bell \& Bower 2000;
 Bell \& de Jong 2000).

Recent observational studies  
have fully supported the view 
(expounded by, e.g., Phillipps et al. 1987 and Impey, Bothun \&
 Malin  1988) 
that low luminosity and low
surface brightness galaxies  numerically dominate the
galaxy population in the local Universe (see e.g. McGaugh 1996; Cross et al. 2001).
However, optical surveys of the local Universe for faint/LSB 
 objects are problematic
 due to the very long exposure times required, the 
  large areas which need to be surveyed and the need to measure
 a redshift for each faint/LSB object found. 
 Consequently, our knowledge of the local population of galaxies 
  at low luminosity and low surface brightness is still relatively limited. 
  This inhibits our knowledge of many broader cosmological/cosmographical
 issues.

Given the limitations of optical surveys for detecting 
 low luminosity / LSB objects, 
 an alternative method to sample the extragalactic population 
 is to use the 21-cm neutral hydrogen (H{\sc I}) line. This provides a way of potentially
 circumventing optical selection effects operating against
  low luminosity and/or LSB objects, since
  a galaxy's H{\sc I} content may be 
  relatively uncorrelated with its optical emission. 
  For example, it is well known that elliptical
 galaxies contain little H{\sc I} whereas we might expect to find 
 large amounts of H{\sc I} in galaxies where star formation has been inefficient, 
 e.g. in low luminosity and LSB galaxies. 
 However, until comparatively recently, most H{\sc I} surveys were limited to 
H{\sc I} measurements of galaxies previously detected in 
optical or IR surveys. 
The advent of the 21-cm multibeam receiving systems 
 at Parkes and Jodrell Bank has made possible, for the first time, 
 blind H{\sc I} surveys of large areas of sky to reasonable sensitivity 
 over comparatively large volumes.  

The H{\sc I} Parkes All Sky Survey (H{\sc I}PASS, Staveley-Smith et al. 1996) 
 was commenced in 1997 and concluded in 2002. H{\sc I}PASS 
 has  surveyed the southern hemisphere (up to $\delta$=+25\degr) to  $cz$=12700~km\,s$^{-1}$  
  and an H{\sc I} mass limit around 
  10$^{6}$d$^{2}_{\rm Mpc}$~M$_{\odot}$.  
  Results from H{\sc I}PASS have indeed added significantly to the census of 
  the local extragalactic 
 population. Recent scientific highlights 
    include the discovery of 10 new members to the Cen A
group (Banks et al. 1999), the detection of an apparently extragalactic H{\sc I}
  cloud with no optical
 counterpart to faint limits (Kilborn et al. 2000) and the discovery of a massive H{\sc I} cloud 
associated with NGC~2442 (Ryder et al. 2001). Kilborn et al. (2002) 
 have recently published a catalogue of 536 galaxies from a 2400 sq deg region of
  H{\sc I}PASS covering the South Celestial Cap. Koribalski et al. (in preparation) will present the 
  H{\sc I}PASS Brightest Galaxies Catalogue (BGC), a 
 catalogue of the brightest 1000 galaxies (in terms of H{\sc I} peak flux) from the whole of H{\sc I}PASS. Meanwhile, Ryan-Weber 
 et al. (2002) have discussed the properties of those previously uncatalogued galaxies 
 found in the BGC.

The H{\sc I} Jodrell All Sky Survey (H{\sc I}JASS) 
 is the northern counterpart to  H{\sc I}PASS. H{\sc I}JASS will survey 
 the northern sky above  $\delta$=22\degr\, to similar 
 sensitivity to H{\sc I}PASS,
 using the Multibeam 4-beam cryogenic receiver mounted on the 76-m Lovell 
 Telescope.  H{\sc I}JASS was begun in 2000. 
 So far  $\simeq$1115~deg$^{2}$ have been surveyed. 
 We recently presented results from the H{\sc I}JASS data covering the M81 group
 (Boyce et al. 2001). 
 The survey reveals several new aspects to the complex morphology of 
 the HI distribution in the group and illustrates that
  a blind HI survey of even such a nearby, well studied 
   group of galaxies can add much new information.

 This paper presents a detailed description of the HI Jodrell 
 All Sky Survey and of the properties of the H{\sc I}-selected  sample of galaxies  
 which has been compiled so far from the H{\sc I}JASS data. We use the sample of confirmed 
 H{\sc I}JASS sources to study 
 the effect that a galaxy's inclination to the line of sight has on its 
 inclusion within an H{\sc I}-selected sample. We show that both highly inclined galaxies 
 and galaxies close to face-on are subject to selection effects which could have 
 led to their being under-represented in previous determinations of the H{\sc I} Mass 
Function (H{\sc I}MF) and H{\sc I} mass density, $\Omega_{\rm HI}$, from H{\sc I}-selected samples of galaxies.

Section 2  describes the hardware, the observing 
 strategy and survey parameters and also describes the data reduction 
 methods. Section 3 describes 
the methods by which galaxies have been detected within H{\sc I}JASS 
 data and their parameters measured. Section 4 is a discussion of the properties 
 of the sample of galaxies found in  H{\sc I}JASS data thus far. In Section 5 we use the sample of confirmed  H{\sc I}JASS 
  sources to study 
 the inclination-dependent selection effects on the inclusion of a galaxy in an H{\sc I}-selected 
 sample and discuss the implications of this. Section 6 presents some concluding remarks.

\section{The Survey}

\subsection{Hardware}

HIJASS uses a cryogenic Multibeam receiver (Bird 1997) of similar design
  to the Multibeam receiver 
 used at Parkes for HIPASS (Staveley-Smith et al. 1996). 
The Multibeam system installed at Jodrell Bank has four dual linearly polarised receivers covering a 
 frequency range  
 of 1200~MHz to 1550~MHz.
The feed horn array consists of 4 stepped circular horns (which 
   were designed  at the CSIRO) arranged in a rhombic pattern, 
 the apertures of which are located at the telescope prime focus. 
  Stepped circular horns were chosen because of their good pattern symmetry, low spillover and good cross polarization properties (Bird 1994). 
  The horns couple directly into a low temperature, 
 high vacuum, cryogenic dewar. 
  The output from the feed horn arrays are then fed to a set of  3-stage 
  high electron mobility transistor (HEMT) pre-amplifiers which are cooled
  to a temperature of around $\sim$25~K. 
    Following amplification,  each receiver 
  RF band is then down converted to an IF bandpass 
  which can be set anywhere between  30~MHz and 245~MHz.

\begin{figure*}
  \epsfig{file=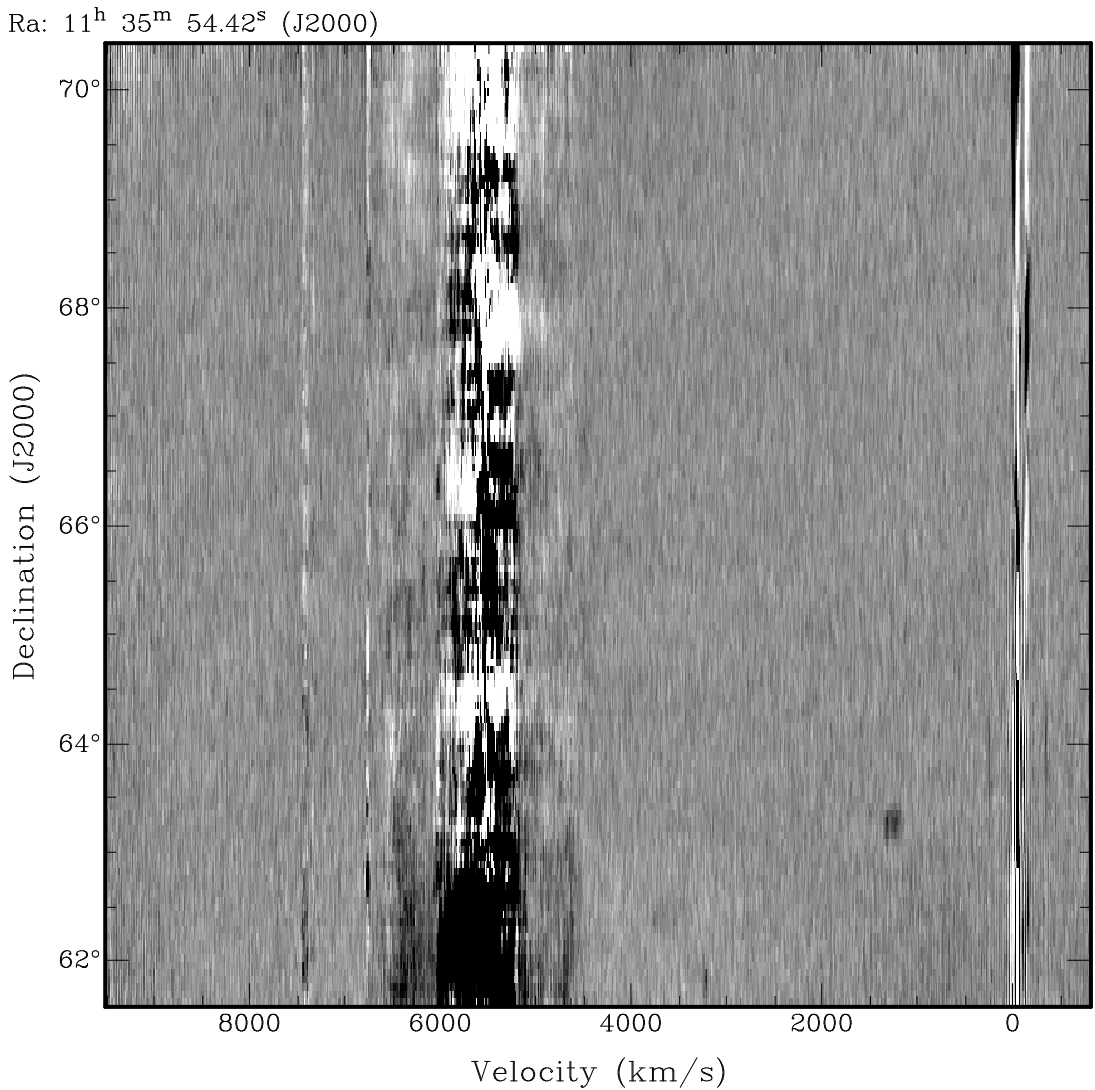}
  \caption{A section of a HIJASS cube at roughly constant $R.A.$. Visible are the residue 
 of Galactic HI at 0~km\.s$^{-1}$, the galaxy UGC06534 at $V_{\odot}\simeq$1300~km\,s$^{-1}$; and the 
 radio frequency interference between 4500$\rightarrow$7500~km\,s$^{-1}$.}
\end{figure*}

 Each of the 8 resultant IF bands are then passed from the focus cabin 
 to the observing room via about 300~m of low loss coaxial cable which 
 is terminated 
 into N-socket connections in the Lovell Observing Room. These outputs 
 are next patched into a set of digitally programmable attenuators 
 and are then fed into a set of equaliser and splitter units.
 Each IF is equalised for frequency dependent cable losses and then 
 split into two outputs. The first output connects to a filter bank which 
 sets the bandpass which is presented to the correlator. 
 A second set of outputs are  used for pulsar survey measurements. 
 
 The correlator was constructed at the Australia Telescope National Facility
   (ATNF). It is built around special purpose VLSI chips 
  developed by the NASA Space Engineering Research Centre for VLSI System Design
 (Canaris 1993). These chips accept 2-bit sampled data streams at rates of up to 
 140 Msamples/sec and form either the cross correlation function of the two streams 
 or the autocorrelation function of one stream at 1025 contiguous sample delays. The chip has 1024 32-bit accumulators and a 32-bit output bus and can integrate for up 
 to 16 seconds. In the multibeam correlator, one of these chips, operating in autocorrelation mode, is used on each of the 8 sampled data streams, thereby providing, after 
Fourier transformation, a measurement of the input spectrum at 1024 contiguous frequencies spaced  at 62.5~kHz
  (equivalent to 13.2~km\,s$^{-1}$ at the rest frequency of HI).

\subsection{Observing strategy and data reduction}

The survey is conducted by  actively scanning the sky 
  in 8\degr\, strips in Declination, at a 
 rate of 1\degr\, per minute. 
 Each declination 
    scan is separated by 10~arcmin 
  but each area of sky is scanned 8 times, resulting in a final scan separation of  1.25~arcmin.  
The data from the 8 correlators are stored every 
 5 s.   A 64~MHz  bandpass with 1024 
 channels  is used, although local interference at the band edges restricts the useful 
  velocity range to about --1000~km\,s$^{-1}$ to +10000~km\,s$^{-1}$ (note, however, that 
  a broad band of radio frequency interference also affects all velocities between 
  4500~km\,s$^{-1}$ and 7500~km\,s$^{-1}$ - see Section 2.3).
  The system temperature is $\sim$30~K.  
   Bandpass correction and calibration  
 are applied using the software package {\sc LIVEDATA}  
   (see Barnes et al. 2001). 
  The spectra are gridded into three-dimensional 
   8\degr$\times$8\degr\,  
 datacubes ($\alpha$,$\delta$,$V_{\odot}$) using the software 
 package {\sc GRIDZILLA} (Barnes et al. 2001).
 The observed spectra are smoothed online by applying a 25 per cent Tukey filter to reduce `ringing' 
 caused by strong Galactic signals entering through the side-lobes. This reduces the actual velocity 
 resolution in the gridded datacubes to 18.1~km\,s$^{-1}$. The spatial 
  pixel size of the datacubes  is 4~arcmin$\times$4~arcmin. 
    The 
 effects of continuum emission on the baselines of the spectra in each cube
 are then removed by the program {\sc POLYCON} written by Daniel Zambonini and Robert Minchin. 
  This program fits and then subtracts a  5th order polynomial baseline to each individual 
  spectrum in a datacube: fitting is only performed on parts of the spectrum free of interference 
 or line emission.

\subsection{Areas Surveyed and Data Quality}

H{\sc I}JASS has been conducted during three observing runs: 
 in April-June 2000; in Jan-Feb 2001; and in Jan 2002.
Table~1 notes those areas of the northern sky so far surveyed by H{\sc I}JASS and during 
 which run the data were taken. The whole of 
 a strip in R.A. between  Decl.=70\degr$\rightarrow$78\degr\, has been surveyed 
(795~deg$^{2}$). Two areas of the R.A. strip between  Decl.=62\degr$\rightarrow$70\degr
   have also been 
surveyed (192~deg$^{2}$), along with smaller areas at Decls.=58\degr, 34\degr, and 26\degr. 
In total 1115~deg$^{2}$ has been surveyed thus far.

\begin{table}
\caption{Areas  surveyed by H{\sc I}JASS}
\begin{tabular}{crrr} \hline
Decl. Range & R.A. Range & Area (deg$^{2}$)  & Run \\ \hline
70\degr--78\degr & complete & 795 & 2000,2001\\ 
62\degr--70\degr & 09$^{h}$02$^{m}$--11$^{h}$55$^{m}$ & 128 & 2001,2002\\
                          & 02$^{h}$30$^{m}$--04$^{h}$02$^{m}$ & 64 & 2002\\
54\degr--62\degr &  03$^{h}$26$^{m}$--04$^{h}$08$^{m}$ & 32 & 2001\\
30\degr--38\degr & 01$^{h}$13$^{m}$--01$^{h}$59$^{m}$ & 64 & 2002\\
22\degr--30\degr & 12$^{h}$08$^{m}$--12$^{h}$34$^{m}$ & 32 & 2002\\ \hline
\end{tabular}
\end{table}

Fig.~1 shows an example of the data. This is a plot of Decl. against Velocity 
 at roughly constant R.A.. 
The H{\sc I}JASS source H{\sc I}JASS~J1133+63 can be seen at Decl.$\simeq$63\degr, $V_{\odot}$=1300~km\,s$^{-1}$. 
This has been identified as  UGC 06534. Prominent in the data is a broad band 
of radio frequency interference (RFI) 
  which affects all velocities from $\simeq$4500~km\,s$^{-1}$ to 
7500~km\,s$^{-1}$. This is mainly due to off-site radio and data communication emissions generated in the 
 locality.

Between the observing runs in 2001 and 2002, the Lovell dish underwent the first stage of 
a major upgrade. A new telescope drive system was installed and around half of the dish 
 surface was replaced.

Prior to this upgrade, the rms noise 
 in the datacubes (away from the broad-band RFI) was typically 16-18~mJy~beam$^{-1}$. 
 However, those cubes made from data taken following the first stage of the refurbishment
 (i.e. in 2002),  
 show an improvement in sensitivity by around 25 per cent, the rms noise in these cubes being 
 about 12-14~mJy~beam$^{-1}$. The improvement is believed to be due to a combination 
 of several factors: the improved 
 sensitivity of the partially resurfaced dish, an improvement in the telescope pointing model, 
  an improvement in the smoothness of the scanning and a concerted effort to reduce local (Jodrell-based)  
  interference during the observing run.

\section{Detecting and Parametrizing Galaxies in H{\sc I}JASS Data}

\subsection{Galaxy detection techniques}

The general method of detecting galaxies in the H{\sc I}JASS data is as follows.
 An initial candidate galaxy list is formed by visually
searching the datacubes. The visual display program {\sc KVIEW} (Gooch 1995) is used 
to search through the data in three dimensions by displaying 2 axes and stepping through the 
third. Narrow velocity-width, bright galaxies are most easily seen when the data are 
displayed in R.A. versus Decl. and stepped through Velocity. Broad velocity-width, faint galaxies 
are more easily discovered when the datacube is displayed R.A. or Decl. versus 
Velocity. The selection criteria for this visual sample is that a detection 
  must : (1) be easily visible above the noise ($\sim$3$\sigma$ in peak flux), 
 (2) should have a spatial extent of greater than 1 pixel, 
 and (3) be visible over two or more velocity planes. Two lists are compiled using 
 these criteria: one of `definite' detections and one of `possible' detections. 
 The list of possible detections includes sources close to the selection limit 
or just below it but still considered possible sources.  

 A second candidate galaxy list is formed by running the automated finding algorithm 
 {\sc POLYFIND}, written by Robert Minchin and Jonathon Davies. {\sc POLYFIND}
  determines the 
 noise in non-masked regions of a hanning smoothed datacube and then looks for 
 peaks at some user-defined 
 level above the noise (typically set at 3$\sigma$). It then runs a series of matched filters 
 over these identified peaks and a peak is noted as a 
 potential source if a sufficiently good fit is obtained.  
The results from {\sc POLYFIND} are then re-checked by eye using the same criteria as 
 used for the eyeball method and lists of definite and possible detections made. 

The two lists of definite detections are then amalgamated and these objects  included 
 in our final sample of confirmed H{\sc I}JASS galaxies. 
The two lists of possible detections are also amalgamated. 
 These objects are then re-observed with the Lovell telescope in single-beam mode 
using a bandwidth of 16~MHz. 
  Those possible sources confirmed by this 
 narrow-band follow-up are then added to the sample of confirmed H{\sc I}JASS sources. 
   The possible sources found during the 2002 run have not yet been subjected to 
 narrow-band follow-up. Hence, for the 2002 data, only the definite detections found by the above 
 selection methods have presently been included in the sample (69 sources).

\begin{table*}
 \epsfig{file=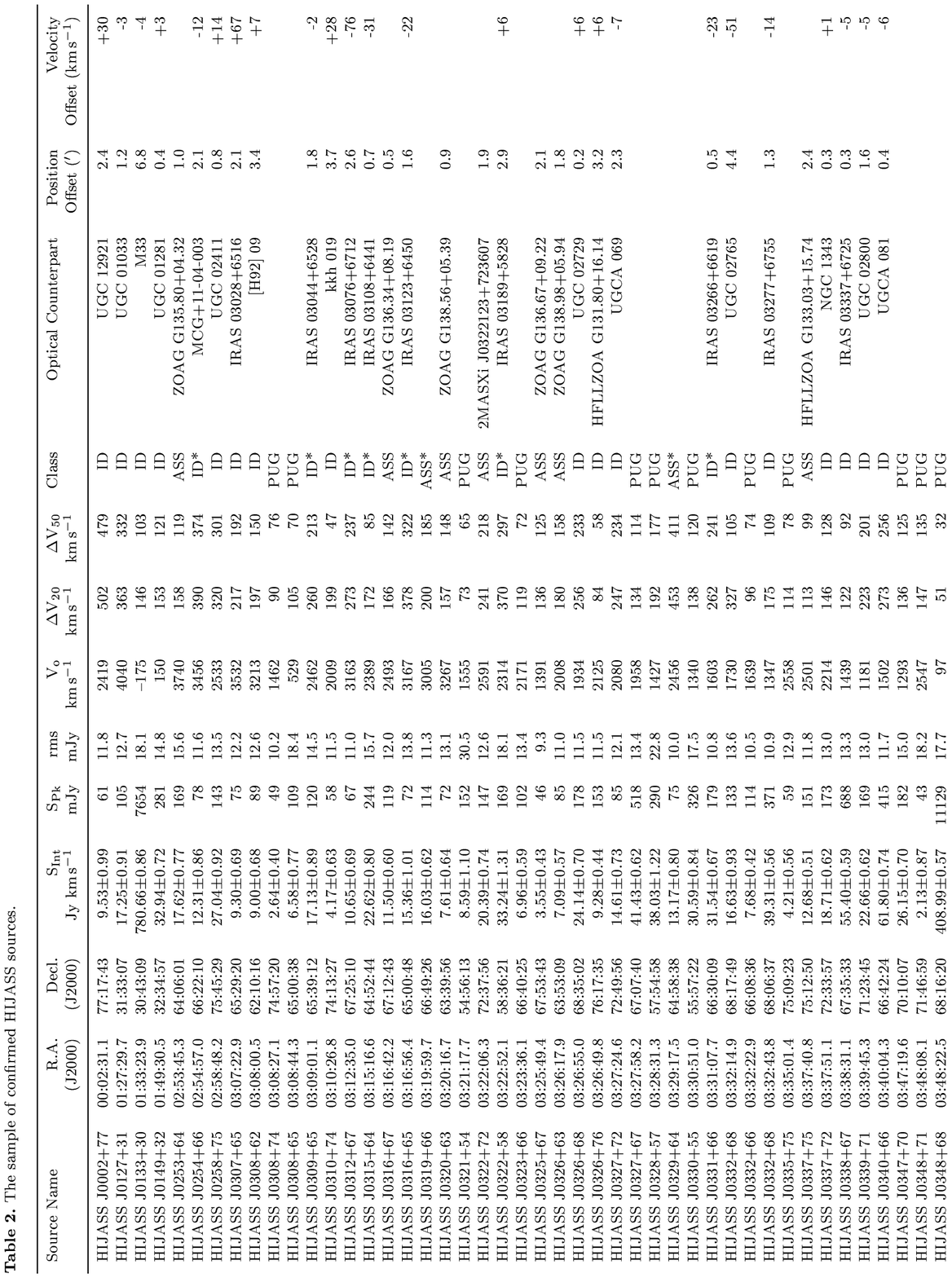}
\end{table*}

\begin{table*}
 \epsfig{file=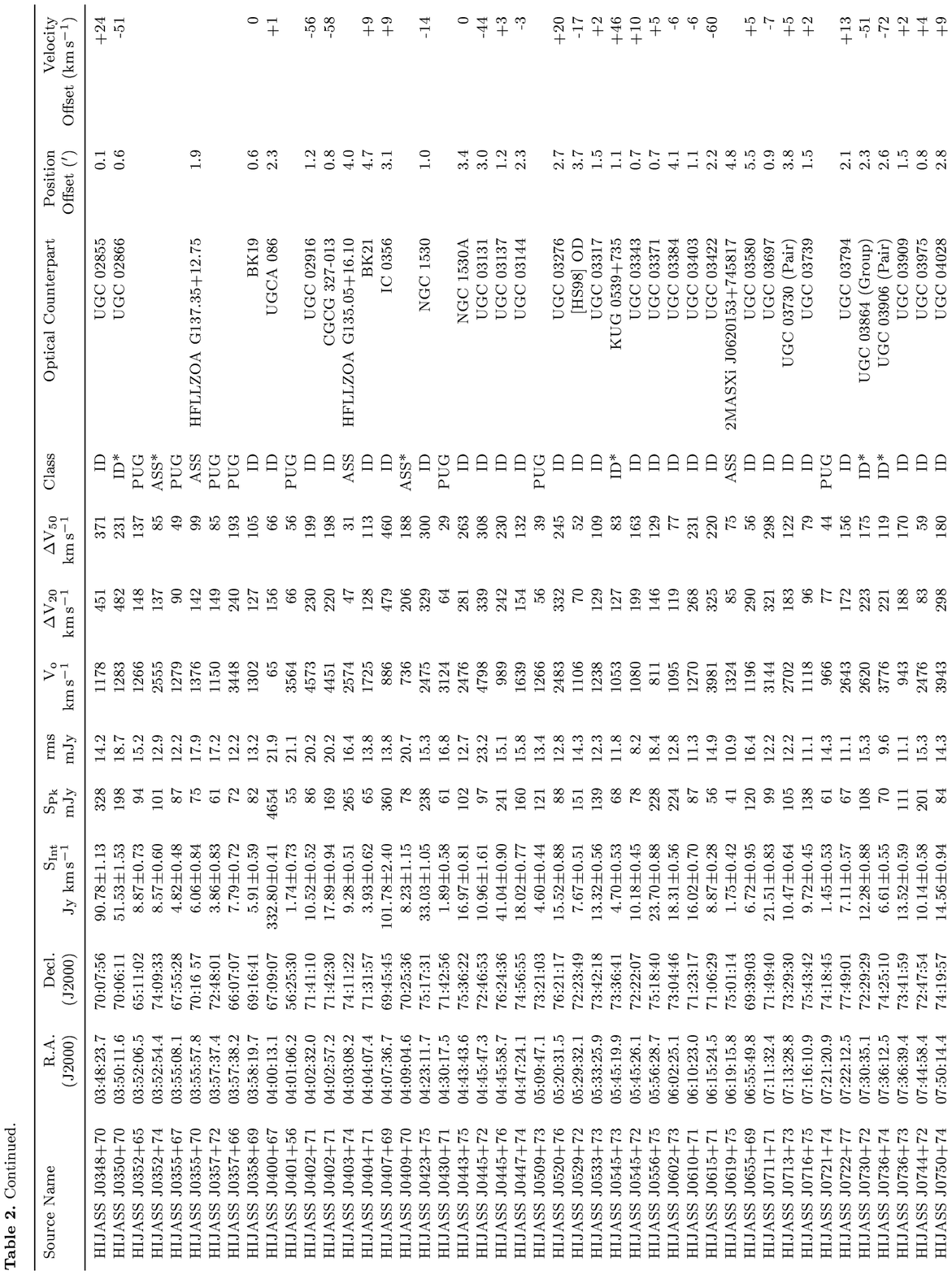}
\end{table*}

\begin{table*}
 \epsfig{file=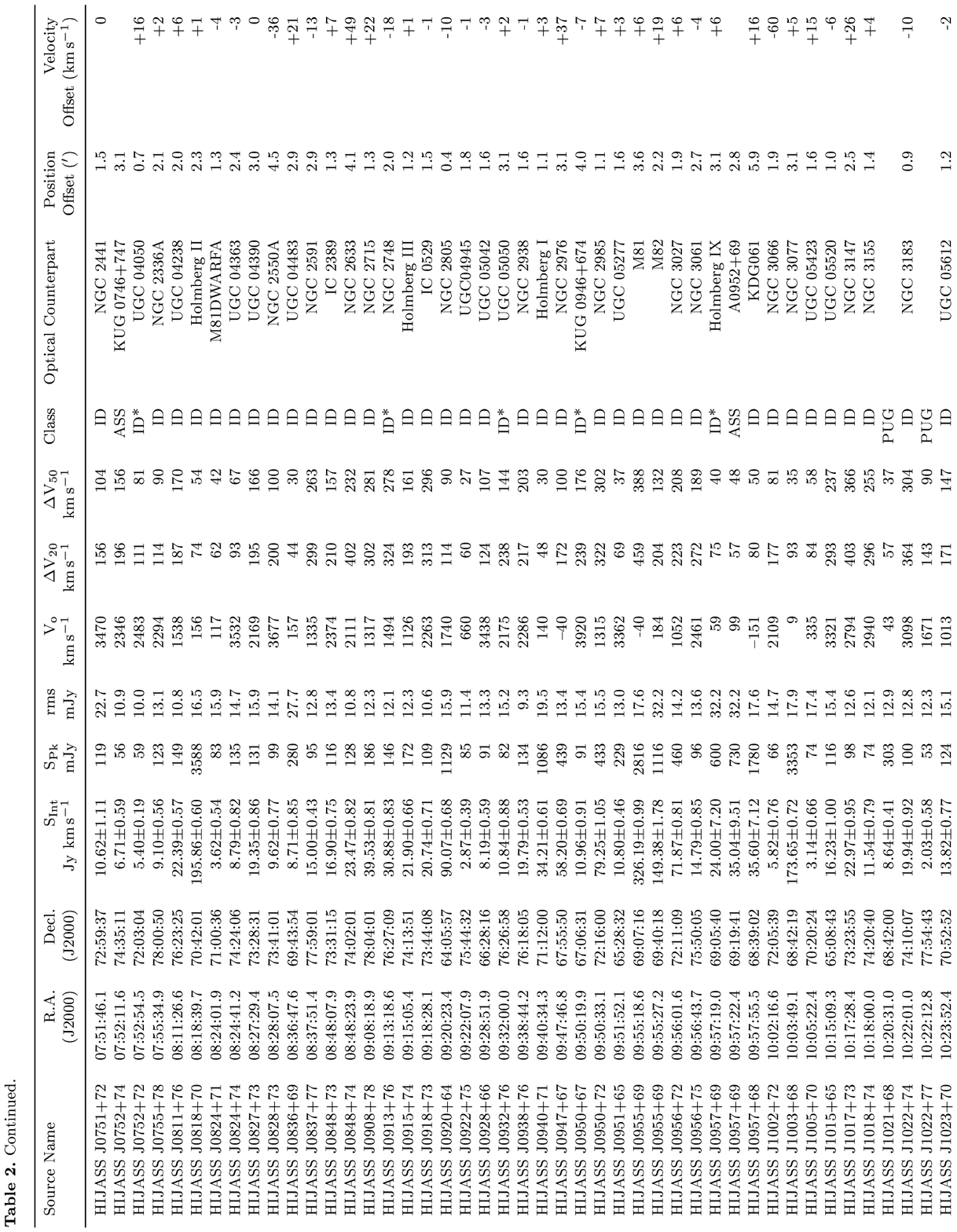}
\end{table*}

\begin{table*}
 \epsfig{file=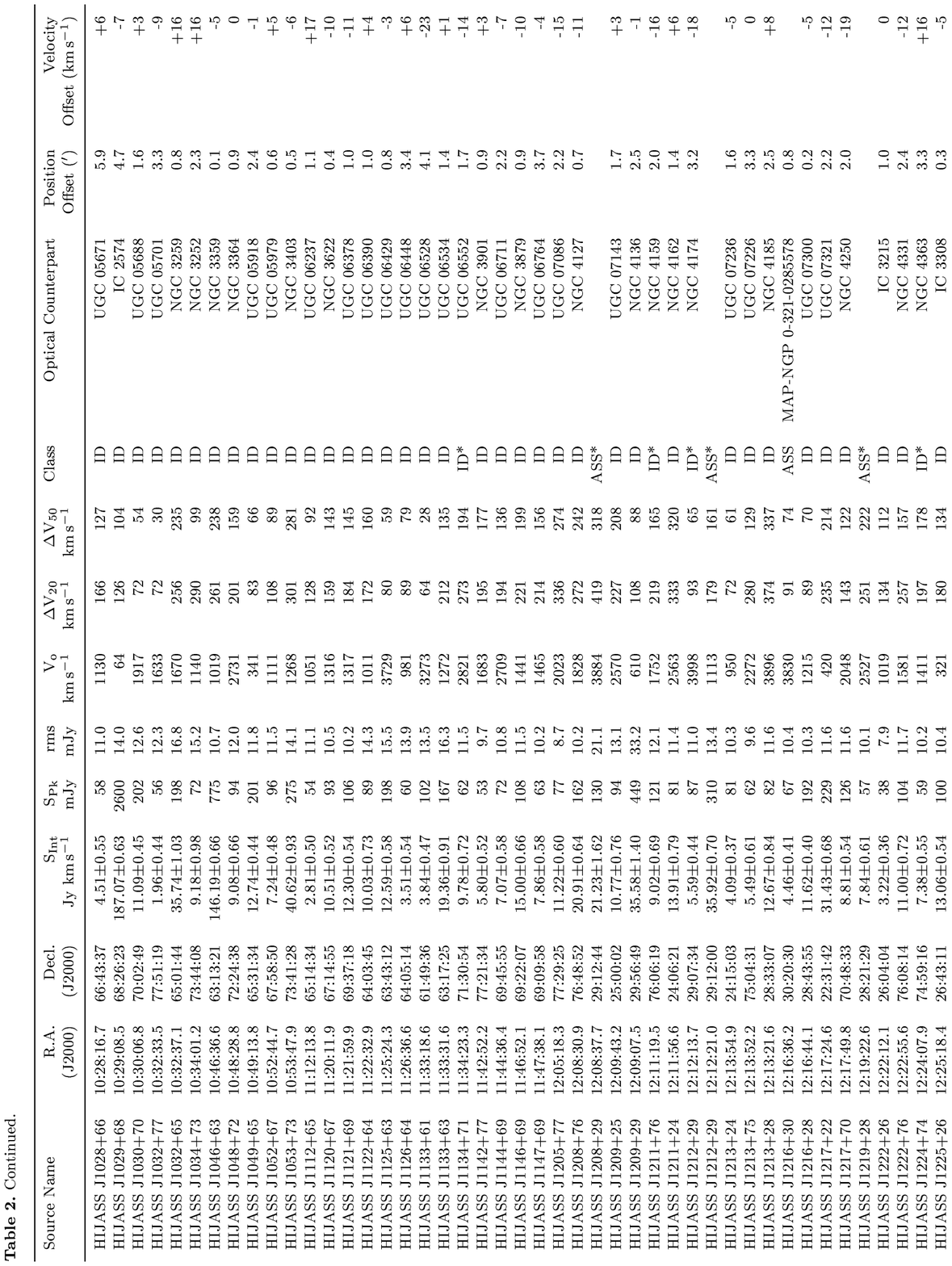}
\end{table*}

\begin{table*}
 \epsfig{file=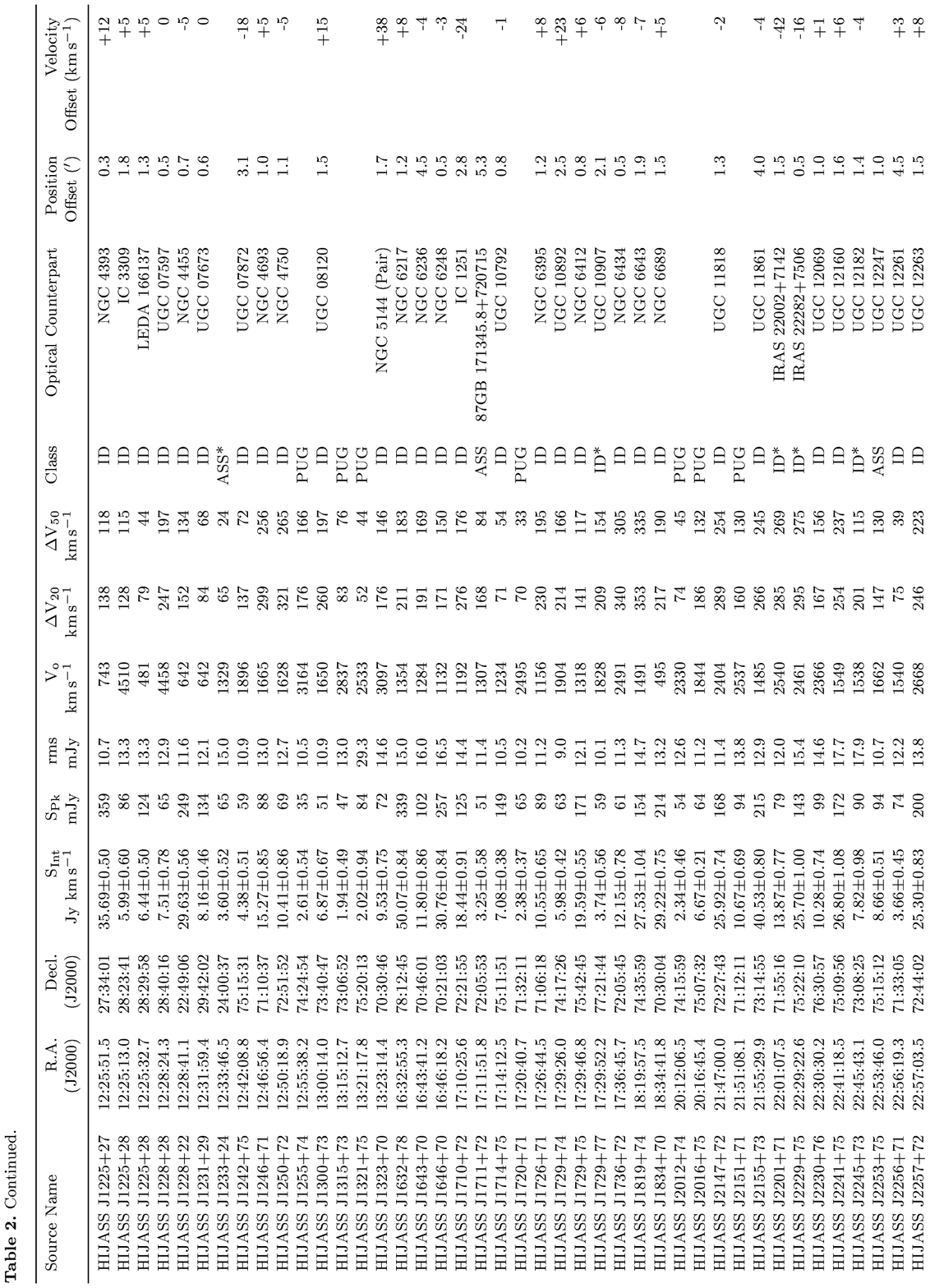}
\end{table*}

\subsection{Parametrization  of the galaxies}

The parameters of those galaxies  included in the sample of confirmed H{\sc I}JASS sources are
   determined from the  data using tasks from the {\sc MIRIAD} software package 
 (Sault et al. 1995). 
Firstly, a two-dimensional Gaussian fit ({\sc IMFIT}) is made to a zeroth order 
 (intensity) moment-map  
    of each detection to determine the central position of the galaxy as well as the 
spatial extent of the H{\sc I}. The central position is then used to generate a spatially 
 integrated spectrum of the detection, using a box size based on the extent of the H{\sc I}. 
 The spectrum is generated using {\sc MBSPECT} which also  gives a measurement of the peak and 
 integrated flux of each detection, as well as the 50 per cent and 20 per cent velocity-width, the rms 
 noise and barycentric velocity.

\begin{figure*}
  \epsfig{file=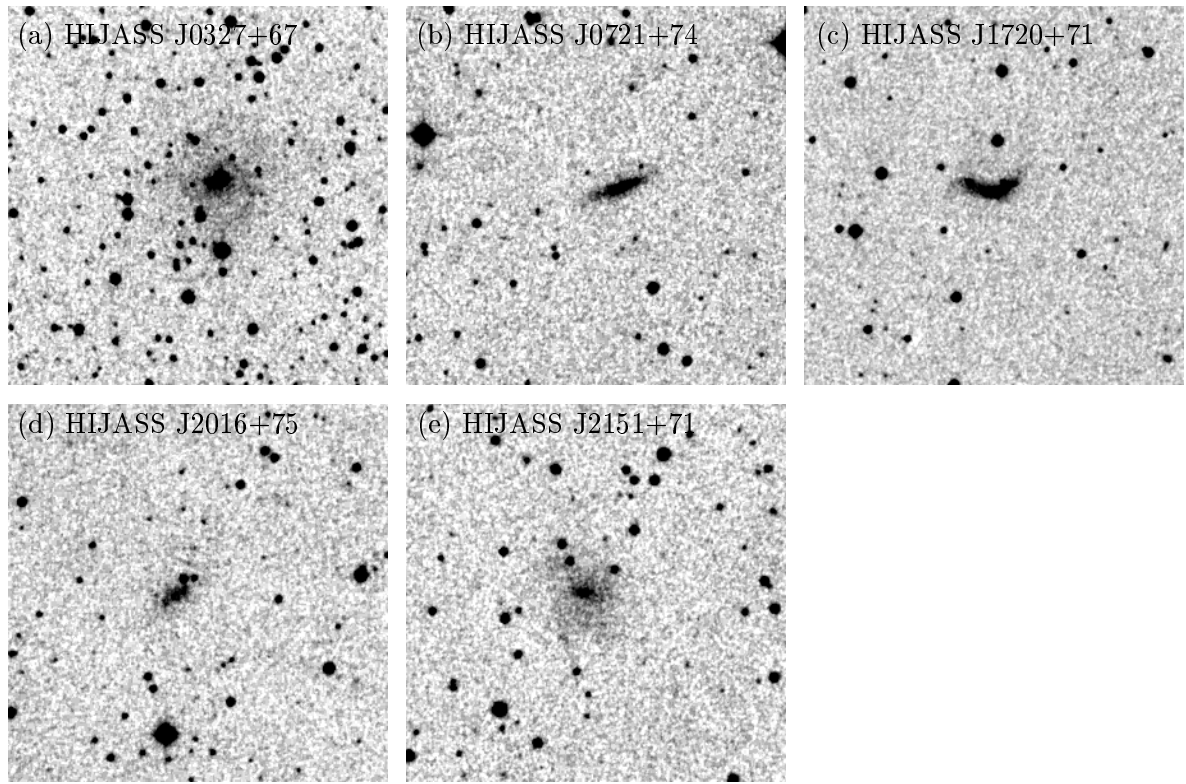}
  \caption{Digitial Sky Survey images of the 5 PUGs for which an obvious and umabiguious optical counterpart can 
 be seen. Each image is 5~arcmin$\times$5~arcmin. }
\end{figure*}

Table~2 presents the derived parameters for the sample of confirmed H{\sc I}JASS sources. 
Column 1 gives the  H{\sc I}JASS  Name.  Columns 2 and 3 give the 
  Right Ascension (J2000) and Declination (J2000) from the {\sc IMFIT} task. 
 Columns 4-6 list  
  the zeroth order moment (Integrated flux S$_{\rm Int}$=$\int S_V dV$), the peak flux 
 ($S_{peak}$), and 
  the noise (rms dispersion around the baseline, $\sigma$);
  all as measured by {\sc MBSPECT}. Columns 7-9 give the first order moment 
 (barycentric velocity V$_{\odot}$), the velocity 
 width at 20 per cent of the peak flux ($\Delta$V$_{20}$), and the velocity width at 
 50 per cent of the peak flux ($\Delta$V$_{50}$), all measured 
 by {\sc MBSPECT} in the radio frame and converted to $cz$. The error in 
 integrated flux is calculated from the rms noise 
 on the spectrum and the velocity extent of the source. 
 
Columns 10-13 contain the details of any counterpart to the 
 H{\sc I}JASS source, as listed in the NASA/IPAC Extragalactic Database (NED).     
 Column 10 contains one of five possible classifications. If there 
 is no object within NED which could be spatially coincident with 
 the H{\sc I}JASS source (defined as being within 
 6~arcmin), then Column 10 contains the classification `PUG' (i.e. Previously Uncatalogued Galaxy). 
  If there is 
 an object in NED which matches the H{\sc I}JASS source in both position and space 
 (defined as being within 6~arcmin and 100~km\,s$^{-1}$) then this is 
 listed as `ID' (Identification).  Those IDs which have been detected in H{\sc I} for the first time by H{\sc I}JASS 
 are denoted by an asterix, i.e. `ID*'. If 
   there is an object within NED which is 
 spatially coincident with the H{\sc I}JASS source (i.e. within 6~arcmin) but for 
 which no redshift is listed in NED, then Column 10 contains the classification 
 `ASS' (i.e. Association). In several cases, there is more than one galaxy 
 within 6~arcmin: in these cases the classification `ASS*' is used. 
 For the ID, ID* and  ASS classifications,  Column 11 lists 
 the object within NED which appears to correspond to the 
 H{\sc I} detection.  Column 12
 lists the position offset (in arcmin) of the coordinates of the 
 optical counterpart  from the HI  position. 
 Column 12 lists (for the IDs and ID*s) the velocity offset of the barycentric velocity 
 contained within NED from the barycentric velocity as measured 
 by H{\sc I}JASS.

\section{Properties of the H{\sc I}JASS galaxies}

\subsection{Composition of the sample}

 There are currently 222 sources included in the sample of confirmed H{\sc I}JASS sources. 
  Of these, 170 (77 per cent) are clearly  associated with a previously 
 catalogued galaxy (classification ID or ID*). However, 25 of these 170 objects have been 
 detected in H{\sc I} for the first time by H{\sc I}JASS (classification ID*). 
 For 4 of these 170 sources, H{\sc I}JASS appears to be measuring H{\sc I} from a pair or a 
 small group of galaxies which lie at the redshift of the H{\sc I}.

There are a further 23 H{\sc I}JASS sources (10 per cent of the whole sample) 
 which lie within 6~arcmin 
 of a catalogued galaxy for which no redshift is reported in 
 NED (classification ASS or ASS*). These H{\sc I}JASS sources may or may not be associated with the 
 catalogued galaxy. 15 of these sources have only one possible optical counterpart within 
 6~arcmin of the H{\sc I} position (classification ASS). We may be relatively confident about the optical 
 identification of these sources. However, even for these 
 sources there remains the possibility that the H{\sc I} has been detected from 
 an associated H{\sc I} cloud (cf. Ryder et al. 2001) or a LSB companion. 
  The other 8 of the 23
  sources have more than one galaxy within 6~arcmin of the H{\sc I}JASS position
 (classification ASS*). We intend to obtain accurate positions for all 
 23 of these sources using H{\sc I} aperture synthesis observations, so as to 
  unambiguously determine the optical counterpart of each source.

 There are then a further 29 sources (13 per cent of the whole sample) which do
 not lie within 6~arcmin of any previously catalogued galaxy (classification PUG). 
 A study of the Digital Sky Survey (DSS) at the positions of these sources reveals an obvious and unambiguous    
  optical counterpart in only 5 cases (J0327+67, J0721+74, J1720+71, J2016+75 and J2151+71). 
  Fig.~2 presents the DSS images of these five  PUGs. 
 Three of these objects (J0327+67, J2016+75, J2151+71) 
 have a  compact but relatively high surface brightness core but 
  a low surface brightness disk. They may have been excluded from optical catalogues because 
 they were mistaken for stars. One object (J0721+74) is a highly inclined 
 but relatively high surface brightness object, although very small ($\simeq$1~arcmin diameter). 
 The fifth object (J1720+71) has a  complex morphology and appears to be 
 involved in some kind of interaction or merger.

 A study of the DSS for the other 
    24 PUGs, reveals no unambiguous optical 
  counterpart. In many cases there are several possible optical candidates within the 
 positional uncertain of H{\sc I}JASS. In several cases, however, no 
 possible  candidate can be seen. The optical counterparts of these sources 
 must be of very low surface brightness. All of the PUGs will 
  have accurate positions determined from 
 aperture synthesis observations and will be the subject of deep optical follow-up 
 work.

It is interesting to  compare the number of PUGs found in the sample of confirmed H{\sc I}JASS sources
   with those 
 found in  the Bright Galaxy 
 Catalogue (BGC: Koribalski et al., in preparation). 
 The BGC contains the 1000 brightest (in H{\sc I} peak flux) sources in the whole of the 
 H{\sc I}PASS sample. 87 of these objects had not been previously catalogued, although 57 of these 
 lie close (within 10$\,^{\circ}$) to the Galactic plane. 
  Of the other 30 previously uncatalogued galaxies within the BGC,  Ryan-Weber et al. (2002) 
    found a single optical counterpart for 25  on the DSS. Whilst the relative 
 number of previously uncatalogued galaxies within the BGC is much smaller 
than within the H{\sc I}JASS sample (3 per cent compared to 13 per cent), most of the BGC 
 objects can be unambiguously assigned to  an optical counterpart on the DSS, whilst 
 most of the H{\sc I}JASS PUGs cannot be.  These differences are probably primarily due to the 
 different flux limits of the two samples. The 
 faintest source in the 
 BGC has a peak flux  of 116 mJy. Only 6 of the 29 H{\sc I}JASS PUGs have a peak flux larger than
 this. The much lower peak flux limit of H{\sc I}JASS has produced a much larger 
 fraction of PUGs compared to the BGC. Since these have generally lower H{\sc I} flux, they 
 are correspondingly harder to detect in optical data.

As it currently stands, 
   the sample of confirmed H{\sc I}JASS sources presents the first H{\sc I} measurement of
  77 galaxies (i.e. classes ID*, ASS, ASS*, PUG), 35 per cent of the whole sample. It 
 presents the first redshift measurement of 52 galaxies (i.e. classes ASS, ASS*, PUG), 
 23 per cent of the whole sample. 
   Between 29 and 52 (13 and 23 per cent)
  of the objects within it have not 
 been previously catalogued. It must also be noted that the 
 `possible' detections from the 2002 observing run have not yet been 
 followed up. Based on the results of previous narrow-band 
 follow-up, this will probably lead to an additional 5-10 sources 
 being added to the sample, many of them previously uncatalogued sources.

\subsection{Peak and integrated flux distributions}

\begin{figure}
  \epsfig{file=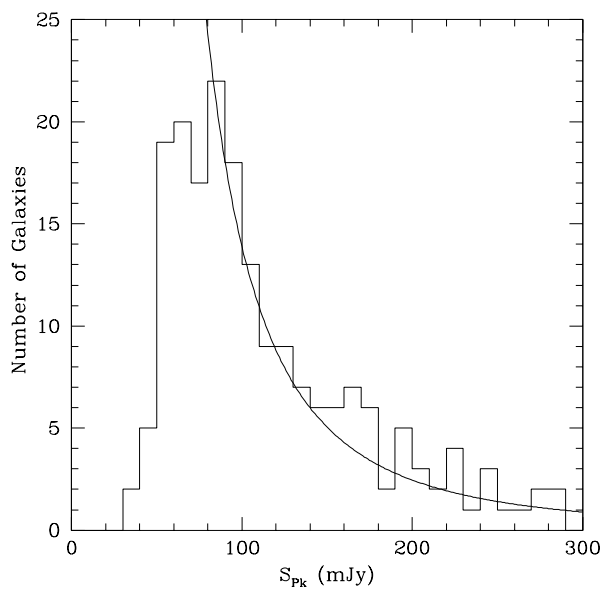}
  \caption{Peak flux distribution of  H{\sc I}JASS sources. The curve shows the best-fit to the 
 data of the 
 function N$_{\rm obj}$$\propto$S$_{\rm Pk}^{-5/2}$, i.e. that expected for a peak flux limited 
  sample of a homogeneous distribution of galaxies.  }
\end{figure}
 
The main factor which determines the 
 inclusion of a galaxy within the H{\sc I}JASS sample ought to be its peak flux
 (rather than integrated flux). 
 Eyeball searches are inevitably  drawn to sources with 
 larger peak fluxes. The {\sc POLYFIND} automated finding algorithm 
 also initially looks for peaks in individual pixels. 
 Fig.~3 is a histogram of the peak flux of every source in the sample of confirmed HIJASS 
 sources.
 For a peak flux limited survey of a homogeneous distribution of galaxies
   we expect N$_{\rm obj}$$\propto$S$_{\rm Pk}^{-5/2}$. The curve on 
 Fig.~3 shows the best fit of this function to the observed  distribution. This implies that our sample is 
  complete to  S$_{\rm Pk}\simeq$80~mJy. 
  We noted above that the rms noise in  H{\sc I}JASS data shows considerable variation 
  between cubes.  
 In particular, the cubes from the 2002 run have considerably lower noise. 
  The completeness limit of 80~mJy  corresponds 
   to a 5$\sigma$ detection in the pre-upgrade data. 
 The 3$\sigma$ detection limit for 
 the post-upgrade data is at $\simeq$39~mJy. Only 2 sources have a peak flux 
 less than this.

  Fig.~4 is a plot of S$_{\rm Pk}$ against 20\% velocity-width ($\Delta$V$_{\rm 20}$). Marked on 
 this is the peak flux detection limit at S$_{\rm Pk}$=39~mJy (3$\sigma$ for the post-upgrade data). 
  This figure also shows another important selection effect in our sample:
  we detect few galaxies at $\Delta$V$_{\rm 20}$$<$50~km\,s$^{-1}$. This is because there is 
 a minimum believable velocity-width which a galaxy must have in order to be selected as a 
 real source from our data. Sources with a narrower velocity-width will be mistaken for 
  narrow band radio frequency interference. From the data it appears that this minimum 
 believable velocity-width is around 4 channels wide for $\Delta$V$_{\rm 20}$, 
 i.e. $\Delta$V$_{\rm 20}^{\rm lim}$=52.8~km\,s$^{-1}$. 
 This makes intuitive sense as it allows 2 `high' channels where the source is seen and believed and 2 `low'
 channels where the flux is dropping off down to the 20 per cent level. 
The locus of this limit is drawn on Fig.~4 
 and constrains the data well. In Section~5 we consider the implications of this velocity-width limit
 for the completeness of HI-selected samples  of galaxies.

\begin{figure}
  \epsfig{file=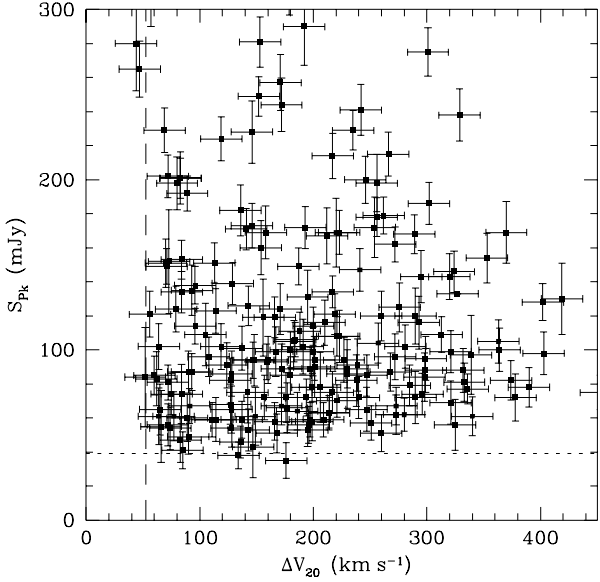}
  \caption{Plot of S$_{\rm Pk}$ against $\Delta$V$_{\rm 20}$ for the sample of confirmed H{\sc I}JASS 
 sources. The short-dashed line shows the 3$\sigma$ peak flux detection limit for the post-upgrade data 
  at 39~mJy. The long-dashed line shows the velocity-width limit equivalent to 4 channels, 
 $\Delta$V$_{\rm 20}^{\rm lim}$=52.8~km\,s$^{-1}$.  }
\end{figure}

 Because the H{\sc I}JASS sample is approximately peak flux limited, 
    the detection of a galaxy of a given integrated flux, S$_{\rm Int}$,   
will (even in similar noise)  be a function of its velocity-width: 
 broader velocity-width 
  galaxies of a given integrated flux have a lower peak flux and therefore 
 are less likely to be detected than a narrower galaxy of the same integrated flux. 
This is clearly seen in Fig.~5, a plot of 
log(S$_{\rm Int}$) against $\Delta$V$_{20}$.
From this can be seen a clear trend for the 
 minimum detected integrated flux to increase with increasing velocity-width.

\begin{figure}
  \epsfig{file=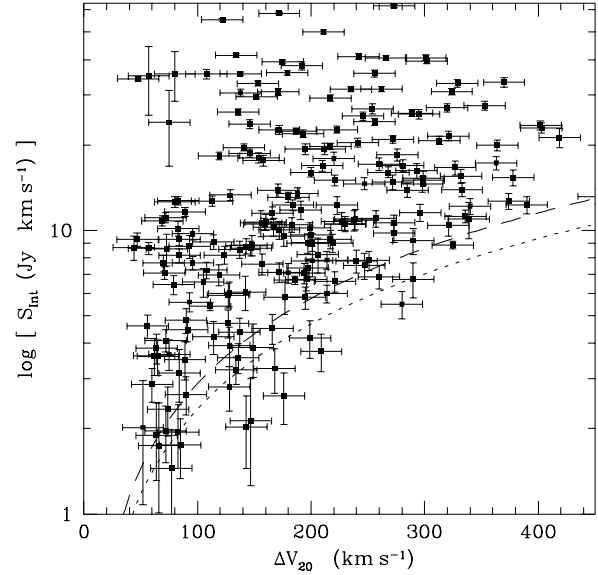}
  \caption{Plot of   log (S$_{\rm Int}$) against  $\Delta$V$_{\rm 20}$ for all the galaxies 
 in the H{\sc I}JASS sample. The long-dashed line is the locus of equation (3) for S$_{\rm pk}^{\rm lim}$=48~mJy 
(i.e. a 3$\sigma$ detection from the pre-upgrade observing runs). The short-dashed line  
   is the locus of equation (3) for S$_{\rm Pk}^{\rm lim}$=39~mJy (i.e. a 3$\sigma$ detection from the post-upgrade observing run). }
\end{figure}

  For a given profile shape we expect 
 the integrated flux, S$_{\rm Int}$, 20\% velocity-width, $\Delta$V$_{\rm 20}$, and the peak flux, S$_{\rm Pk}$, to be related via 
\begin{equation}
S_{\rm Int} = k \, \Delta V_{20} \,  S_{\rm Pk}
\end{equation}
where k is a constant which depends on the profile shape. For a top-hat function k$\simeq$1, for a 
  Gaussian k$\simeq$0.7. Fig.~6 is a plot of S$_{\rm Int}$ against $\Delta$V$_{\rm 20}$.S$_{\rm Pk}$ for all 
 galaxies in the H{\sc I}JASS sample. The linearity of this relationship is clear although there is some 
 expected scatter in the value of k. A least-squares best-fit to this data produces a mean value of k$\simeq$0.6. Adopting this 
 value we can say
\begin{equation}
S_{\rm Int} \simeq 0.6 \, \Delta V_{\rm 20} \, S_{\rm Pk}
\end{equation}
for the H{\sc I}JASS sample. Hence, for a given peak flux limit, S$_{\rm Pk}^{\rm lim}$, 
 the integrated flux limit, S$_{\rm Int}^{\rm lim}$ is a function of $\Delta$V$_{\rm 20}$ via:   
\begin{equation}
S_{\rm Int}^{\rm lim} \simeq 0.6 \, \Delta V_{\rm 20} \, S_{\rm Pk}^{\rm lim}
\end{equation}
 The loci of this relationship for  S$_{\rm Pk}^{\rm lim}$=48~mJy 
(i.e. a 3$\sigma$ detection from the pre-upgrade observing runs) 
 and for S$_{\rm Pk}^{\rm lim}$=39~mJy (i.e. a 3$\sigma$ detection from the post-upgrade observing run) are plotted 
 on Fig.~5. 
 These describe well the form of the observed cut-off in S$_{\rm Int}$ as a function of $\Delta$V$_{\rm 20}$.

\begin{figure}
  \epsfig{file=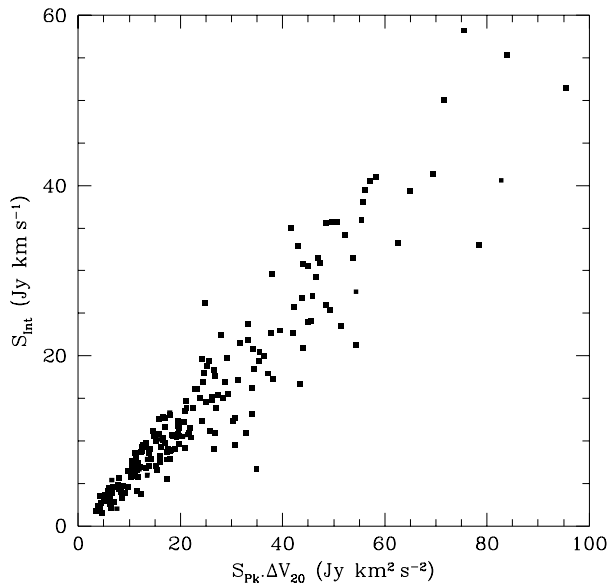}
  \caption{Plot of  S$_{\rm Int}$ against $\Delta$V$_{\rm 20}$.S$_{\rm Pk}$ for the galaxies in the sample of 
 confirmed H{\sc I}JASS sources. }
\end{figure}

It is worth noting that the fact that our sample is essentially peak flux limited has a dramatic effect 
 on the proportion of broad velocity-width to narrow velocity-width galaxies 
  included in the sample, compared to the proportion we would expect to find in an integrated flux 
 limited sample.
For the sake of illustration, we consider the fraction of galaxies with S$_{\rm Int}>$3~Jy~km\,s$^{-1}$ 
 which will be included in the H{\sc I}JASS sample as a function of velocity-width. The value of 
  3~Jy~km\,s$^{-1}$ corresponds to a galaxy of velocity-width 100~km\,s$^{-1}$ with a peak flux of 
 about 48~mJy (the 3$\sigma$ limit for the pre-upgrade data).   At $\Delta$V$_{\rm 20}$=100~km\,s$^{-1}$, 
 all galaxies with  S$_{\rm Int}>$3~Jy~km\,s$^{-1}$ will be included in the HIJASS sample. However, 
 at broader velocity-widths, the integrated flux limit will increase [equation (3)] and the sample will contain  a progressively
 smaller fraction of galaxies with S$_{\rm Int}>$3~Jy~km\,s$^{-1}$.

 In Table~3 we list (Column 2) the S$_{\rm Int}^{\rm lim}$ values equivalent to a range of $\Delta$V$_{\rm 20}$ values (Column 1)
 using equation (3) and assuming  S$_{\rm Pk}$=48~mJy. For each pair of 
 $\Delta$V$_{\rm 20}$, S$_{\rm Int}^{\rm lim}$ values we list (Column 3) 
 the fraction of galaxies with S$_{Int}>$3~Jy~km\,s$^{-1}$ which will be missing from the HIJASS sample. This 
 has been calculated assuming a homogeneous distribution of sources with N$_{\rm obj}$$\propto$S$_{\rm Int}^{-5/2}$ at 
 each $\Delta$V$_{\rm 20}$.

\setcounter{table}{2}
\begin{table}
\caption{The fraction of galaxies with S$_{\rm Int}>$3 Jy~km\,s$^{-1}$ which will be missed from the H{\sc I}JASS sample 
 as a function of $\Delta$V$_{\rm 20}$. This assumes S$_{\rm Pk}^{\rm lim}$=48~mJy and that the distribution 
 of galaxies in space at each $\Delta$V$_{\rm 20}$ is homogeneous.}
\begin{tabular}{lrrr} \hline
 $\Delta$V$_{\rm 20}$ & S$_{\rm Int}^{\rm lim}$ & frac of  \\ 
 km\,s$^{-1}$  & Jy km\,s$^{-1}$ &  gals missed \\  \hline
150 & 4.32 & 0.42 \\
200 & 5.76 & 0.62 \\
250 & 7.20 & 0.73 \\
300 & 8.64 & 0.80 \\
350 & 10.08 & 0.84 \\
400 & 11.52 & 0.87 \\ \hline
\end{tabular}
\end{table}

\begin{figure}
  \epsfig{file=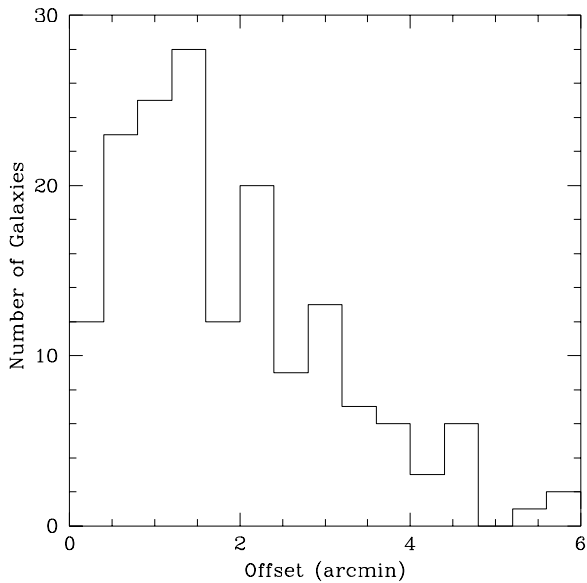}
  \caption{Histogram showing the distribution of the offset between H{\sc I}JASS source position 
 and the position of the identified optical counterpart. }
\end{figure}

 At $\Delta$V$_{\rm 20}$=150~km\,s$^{-1}$ only 68 per cent of galaxies with S$_{\rm Int}$$>$3~Jy~km\,s$^{-1}$ will be included in the HIJASS sample. 
   At $\Delta$V$_{\rm 20}>$300~km\,s$^{-1}$ less than  20 per cent of galaxies with S$_{\rm Int}$$>$3~Jy~km\,s$^{-1}$ will be included. This is a particularly important selection effect since we might reasonable expect that broader velocity-width 
 galaxies will tend to have higher HI masses (see e.g. Rao \& Briggs 1993). Hence, compared to an integrated flux limited sample, 
 our selection techniques may be significantly biased against the inclusion of  higher H{\sc I} mass galaxies. This effect will not bias an  
  H{\sc I}MF derived from the data provided that the selection effect is properly accounted for. 
  It does, however, mean that the morphological mix of galaxies  revealed by a blind H{\sc I} survey is going to be biased towards 
  narrow velocity-width dwarf galaxies and away from  broad velocity-width giant galaxies. This bias needs to be born in mind
  when considering the relative proportions of the different morphologies of galaxies in an H{\sc I}-selected sample.

\subsection{Positional accuracy of H{\sc I}JASS}

\begin{figure}
  \epsfig{file=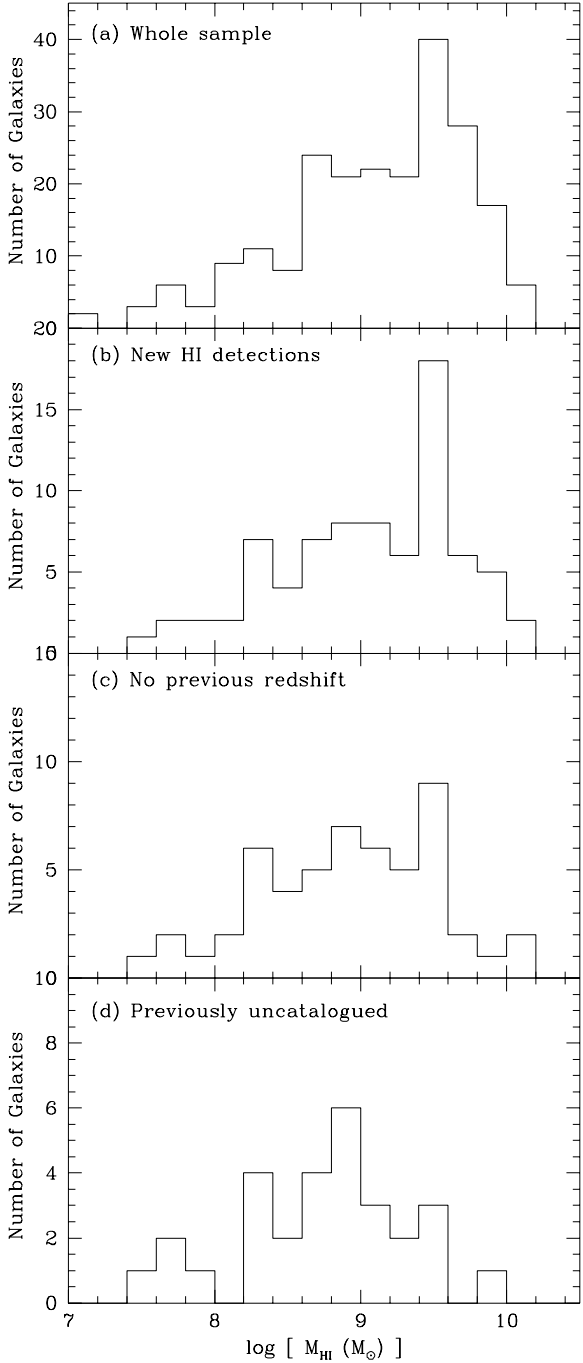}
  \caption{(a) M$_{\rm HI}$ distribution of the whole of the H{\sc I}JASS sample; (b) M$_{\rm HI}$ distribution of the 77 
 galaxies which had not previously been detected in H{\sc I} (classes ID*, ASS, ASS*, PUG); (c) M$_{\rm HI}$ distribution of 
 the 52 galaxies with no previous redshift measurement  (classes ASS, ASS*, PUG);
 (d) M$_{\rm HI}$ distribution for the 29 PUGs.
 }
\end{figure}

\begin{figure}
  \epsfig{file=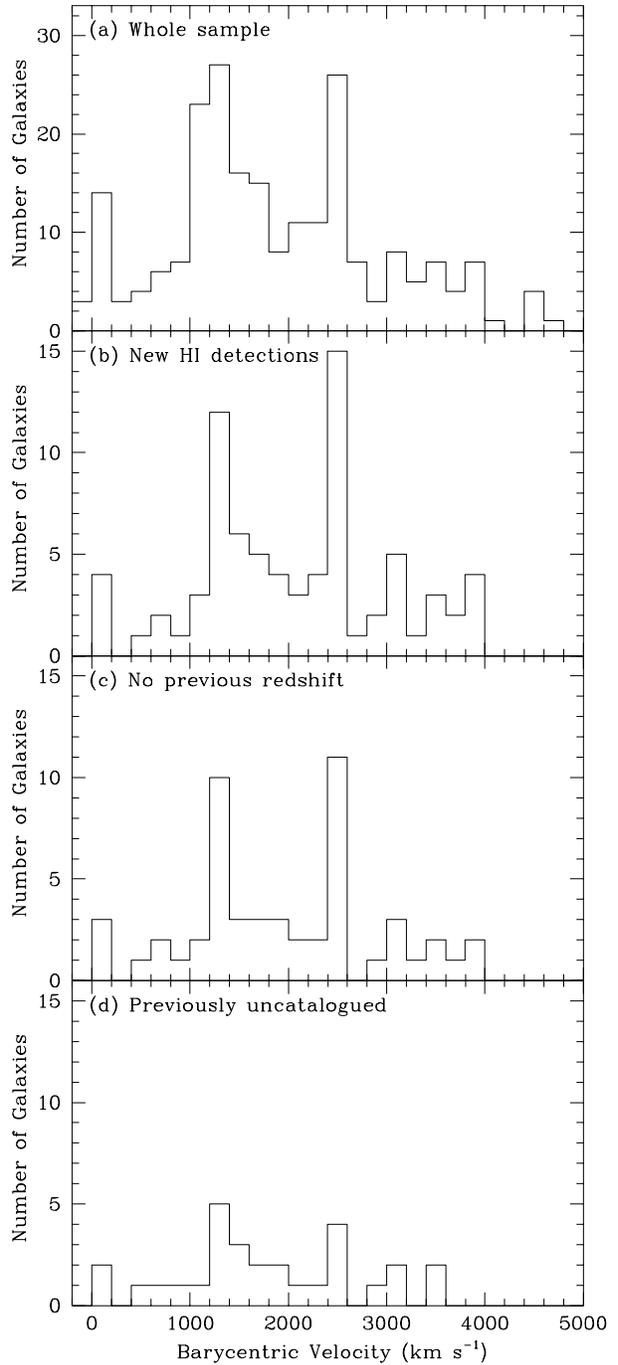}
  \caption{(a) The $cz$ distribution of the whole of the H{\sc I}JASS sample; (b) the $cz$ distribution of the 77
 galaxies which had not previously been detected in H{\sc I} (classes ID*, ASS, ASS*, PUG); (c) the $cz$ distribution of 
 the 52 galaxies with no previous redshift measurement (classes ASS, ASS*, PUG);
 (d) the $cz$ distribution for the 29 PUGs.}
\end{figure}

The positional accuracy of H{\sc I}JASS sources can be judged by 
 considering the offset between the H{\sc I}JASS positions and the 
 positions listed in NED for those galaxies identified as being 
 associated with each H{\sc I}JASS source (i.e. the IDs and ID*s).
  Fig.~7 shows a histogram of these offsets. 
 The majority of H{\sc I}JASS sources  (71 per cent) lie within 2.5~arcmin of the 
  NED position, with only  a very small fraction (7 per cent) lying 
 beyond 4~arcmin.

\subsection{Mass distribution}

Fig.~8(a) shows a histogram of the distribution of H{\sc I} masses 
 for the whole of the sample of confirmed H{\sc I}JASS sources. Galaxies within 
 the M81 group have been assumed to lie at 3.63~Mpc (Freedman et al. 1994). 
 The distances of the other galaxies have been determined from their redshifts 
 (assuming H$_{o}$=75~km\,s$^{-1}$~Mpc$^{-1}$). 
   There is a peak in this distribution at 
  $\sim$10$^{9.6}$~M$_{\odot}$. This is close to the value found for 
 $M_{\rm HI}^\star$ of 10$^{9.75}$~M$_{\odot}$ by  Zwaan et al. (1997).

Although the H{\sc I}JASS bandpass stretches to 10000~km\,s$^{-1}$,  
   the RFI problem beyond $cz$=4500~km\,s$^{-1}$ effectively places a bandpass 
limit at this point. A galaxy with an H{\sc I} mass of 10$^{9.75}$~M$_{\odot}$ 
  would have an integrated flux of about 6.6~Jy~km\,s$^{-1}$ at this $cz$=4500~km\,s$^{-1}$. 
  This is similar to the limiting integrated flux one would expect for a broad 
velocity-width ($\ga$200~km\,s$^{-1}$) galaxy. Hence, H{\sc I}JASS is effectively 
   bandpass-limited for galaxies 
 with $M_{\rm HI}$$\ga$$M_{\rm HI}^{\star}$ (assuming Zwaan et al.'s value for 
  $M_{\rm HI}^{\star}$). For galaxies with  $M_{\rm HI}$$\la$$M_{\rm HI}$$^{\star}$, 
 H{\sc I}JASS is flux limited.

Fig.~8(b) shows the H{\sc I} mass distribution only for those 77 galaxies which 
 had not previously been detected in H{\sc I} (i.e. classes ID*, ASS, ASS* and PUG) . This distribution is very similar
  to that of the whole sample. 
 In fact, most of those 25 previously catalogued galaxies which had not 
 previously been detected in H{\sc I}, have H{\sc I} masses 
 between 10$^{9.3}$--10$^{9.8}$~M$_{\odot}$. All but 2 have integrated 
 fluxes above 10~Jy~km\,s$^{-1}$. 
 These would appear to mostly 
 be `normal' galaxies which had simply not been observed in H{\sc I} 
 prior to H{\sc I}JASS.

Fig.~8(c) shows the H{\sc I} mass distribution for the 52 H{\sc I}JASS objects which had no previous redshift 
 measurement (classes ASS, ASS*, PUG).  
 The peak at 10$^{9.5}$~M$_{\odot}$ is much less pronounced in 
 this plot. A weaker peak in this  distribution can be seen at 10$^{8.9}$~M$_{\odot}$.  
  A peak at this mass is clearly seen 
  in Fig.~8(d) which plots the H{\sc I} mass distribution  for the 29 PUGs. 
 The peak in this distribution is at about 10$^{8.9}$~M$_{\odot}$, an 
 order of magnitude below the peak in the distribution for the 
 whole sample. 
 In fact  69 per cent of the PUGs 
  lie at $M_{\rm HI}$$<$10$^{9}$~M$_{\odot}$. In comparison 
   only 39 per cent of the full sample lie in this mass range. 
 A similar result was found by Ryan-Weber et al. (2002) for the 
 30 previously uncatalogued galaxies in the BGC at $|$b$|$$>$10\degr. 
 They found the mass distribution of these 30 galaxies to peak at 
 10$^{8.7}$~M$_{\odot}$, compared to a peak at 10$^{9.5}$~M$_{\odot}$ for the 
 whole of the BGC.    
 This tendency for the newly discovered galaxies to have lower H{\sc I} masses  
  suggests a correlation between H{\sc I} mass and optical detectability.

\begin{figure*}
  \epsfig{file=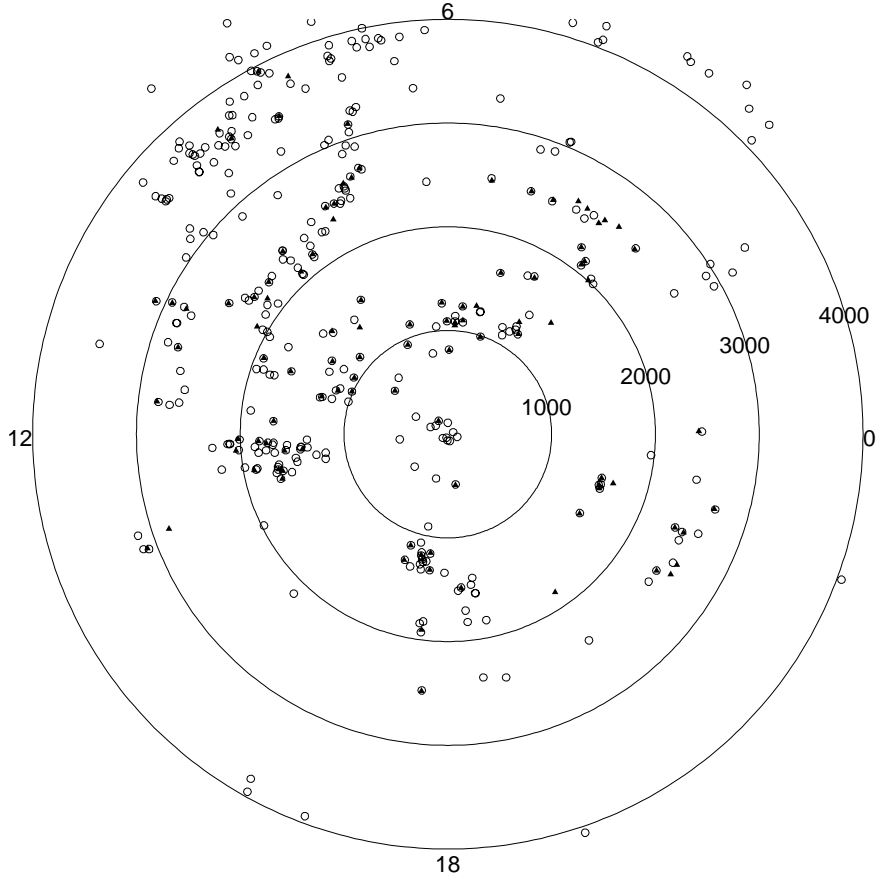}
  \caption{Distribution of R.A. vs $cz$ of H{\sc I}JASS galaxies in the range Decl.=70\degr$\rightarrow$78\degr (filled triangles). Also shown (open circles) is the R.A. vs $cz$ distribution of all galaxies within 
 NED within this declination range and with measured redshifts.}
\end{figure*}

\subsection{Velocity distribution}

Fig.~9(a) shows the distribution of all the H{\sc I}JASS galaxies  as a
 function of $cz$. The peak close to 0~km\,s$^{-1}$ is due to galaxies 
 in the Local Group and the M81 Group. Beyond that there are prominent
peaks at $cz$$\sim$1200~km\,s$^{-1}$ and 2500~km\,s$^{-1}$. 
 These are due to large-scale structure. 
 Note that only a handful of galaxies have been identified beyond 4500~km\,s$^{-1}$. 
  This is mainly due to the increasing problems of RFI beyond this point 
 and the difficulty of distinguishing any galaxies from interference.

Fig.~9(b) shows the $cz$ distribution for all of those 77 galaxies which 
 had not previously been detected in H{\sc I} (Classes ID*, ASS, ASS*, PUG). 
  This distribution is very similar
  to that of the whole sample.  Fig.~9(c) shows the $cz$ 
  distribution for the 52 objects which had no previous redshift 
 measurement (classes ASS, ASS*, PUG).
 Fig.~9(d) plots the H{\sc I} mass distribution just for the 29 PUGs. All 
  of these distributions show the same 
 peaks at $cz$=1200~km\,s$^{-1}$ and 2500~km\,s$^{-1}$ as that seen in Fig~8(a). 
 Clearly, the distribution of new H{\sc I} detections, new redshift measurements and 
 newly catalogued galaxies follows that of the large-scale structure as revealed 
 by the whole sample. This is particularly interesting in regard to the 
 previously uncatalogued galaxies. These have a H{\sc I} mass distribution with a
peak an order of magnitude lower than that of the whole sample (see Fig.~8(d)) but 
 the $cz$ distribution is not skewed to nearer distances.

The relationship of  H{\sc I}JASS galaxies to large-scale structure is further 
 explored in Fig.~10. This is a  diagram showing the relationship of H{\sc I}JASS 
 galaxies (filled triangles) and all objects with 
  redshifts in NED (open circles) in the complete $R.A.$ 
 strip between $Decl.$=70\degr$\rightarrow$78\degr. The general association of H{\sc I}JASS 
 sources with the large-scale structure as delineated by the NED objects 
  is clear. Most of those 
 previously uncatalogued objects found by H{\sc I}JASS lie in regions already populated 
 with galaxies. Some of the structures which can be seen in the data 
 include the group of galaxies at R.A.$\sim$17$^{h}$, $V_{\odot}$$\sim$1200~km\,s$^{-1}$ 
 (which includes NGC~6217, NGC~6236, NGC~6248 and NGC~6395); the NGC~4291 group at 
   R.A.$\sim$13$^{h}$, $V_{\odot}$$\sim$1600~km\,s$^{-1}$; a group including 
  UGC~03317, UGC~03343 and UGC~03403 at  R.A.$\sim$6$^{h}$, $V_{\odot}$$\sim$1200~km\,s$^{-1}$; 
  and two prominent `walls', at R.A.$\sim$3$^{h}$$\rightarrow$5$^{h}$, $V_{\odot}$$\sim$2600~km\,s$^{-1}$ 
  and at R.A.$\sim$7$^{h}$$\rightarrow$10$^{h}$, $V_{\odot}$$\sim$2300~km\,s$^{-1}$.

We have thus far failed to positively detect any galaxy beyond 
 $cz$=4800~km\,s$^{-1}$. The presence of RFI between $cz$=4500--7500~km\,s$^{-1}$ 
 makes the detection of galaxies in the region practically impossible. 
  None the less, the band between 7500-9000~km\,s$^{-1}$ is generally 
  free from RFI and we ought to be able to detect any galaxies
 in this region. However,  to be detectable beyond $cz$=7500~km\,s$^{-1}$, 
 a galaxy would need an H{\sc I} mass of $\ga$10$^{10.9}$~M$_{\odot}$. 
 Such objects must be very rare. For example, Kilborn et al.'s (2002)
 survey of 2400~deg$^{2}$ of H{\sc I}PASS data did not detect any galaxy 
 this massive. Minchin (2001) presented results from surveying a 
 32~deg$^{2}$ with the Parkes multibeam system to 12$\times$ 
 the standard H{\sc I}PASS exposure time.  No galaxy with $M_{\rm HI}>$10$^{10.6}$~M$_{\odot}$ 
 was detected.

\subsection{Comparison to H{\sc I}PASS}

 Following  the first stage of the Lovell telescope upgrade, 
 the rms noise in H{\sc I}JASS data is around 12-14~mJy~beam$^{-1}$. 
  The typical noise in H{\sc I}PASS data is $\sim$14~mJy~beam$^{-1}$, 
 although there is considerable variation between H{\sc I}PASS cubes 
 with rms spanning the range  9-17~mJy~beam$^{-1}$.

The data at R.A.$\simeq$12$^{h}$08$^{m}$$\rightarrow$12$^{h}$34$^{m}$,  
 Decl.=22\degr$\rightarrow$30\degr, observed during the 2002 run, was taken 
 specifically because it provides a small overlap with the H{\sc I}PASS
 survey between Decl.=22\degr$\rightarrow$25\degr. 
Fig.~11 shows the integrated fluxes from H{\sc I}PASS and H{\sc I}JASS cubes for those 
  galaxies which lie in the overlap region. The calibration between 
 the surveys appears robust.

As noted in Section~4.4, 
 terrestrial-based RFI 
 effectively limits the H{\sc I}JASS bandpass to $cz$$<$4500~km\,s$^{-1}$ whereas 
 H{\sc I}PASS can survey out to 12700~km\,s$^{-1}$.
 However, only galaxies 
 with  $M_{\rm HI}\ga$M$_{\rm HI}^{\star}$ can be seen beyond 4500~km\,s$^{-1}$ in H{\sc I}PASS 
 data. In fact less than 24 per cent of the H{\sc I}PASS sample presented by Kilborn et al. (2002)
   lies beyond 4500~km\,s$^{-1}$. H{\sc I}PASS therefore, has an advantage over H{\sc I}JASS in that 
  very massive sources can be detected over larger volumes. However, for galaxies 
 with  $M_{\rm HI}\la$$M_{\rm HI}^{\star}$, both H{\sc I}JASS and H{\sc I}PASS are flux limited 
 and H{\sc I}JASS is potentially  more sensitive.

\section{Inclination-Dependent Selection Effects in an H{\sc I}-Selected Sample}

The distribution function of neutral hydrogen masses among galaxies and intergalactic clouds (the H{\sc I} mass function, H{\sc I}MF) 
 and, more generally, the neutral hydrogen density in the local Universe, $\Omega_{\rm HI}$, are important inputs 
 into models of cosmology and galaxy evolution. 
 Prior to the advent of blind H{\sc I} surveys, astronomers were restricted to constructing an H{\sc I}MF by making H{\sc I} measurements 
 of optically selected samples of galaxies (see, e.g. Rao \& Briggs 1993; Solanes, Giovanelli \& Haynes 1996).

\begin{figure}
  \epsfig{file=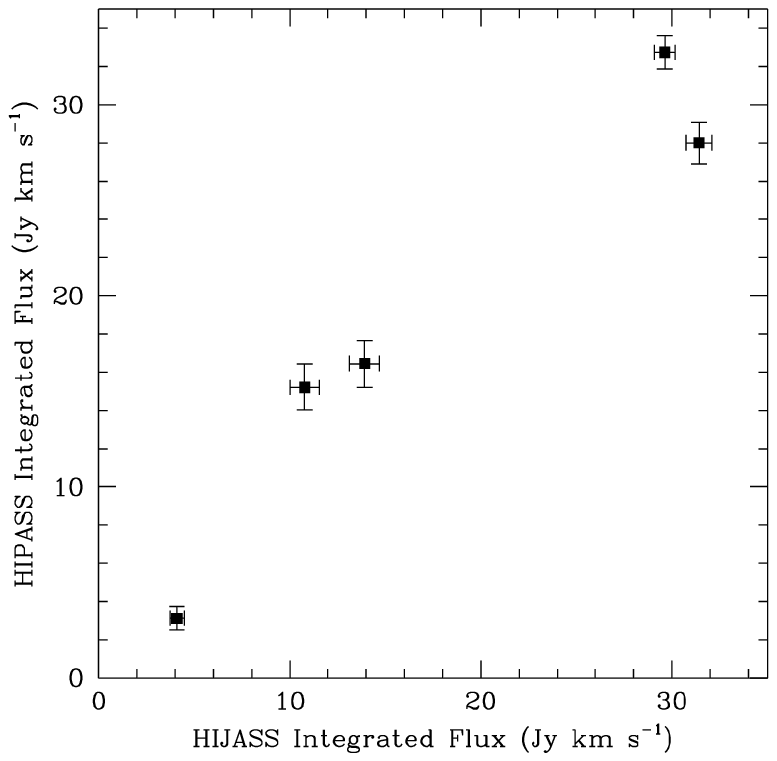}
  \caption{Comparison of integrated fluxes from H{\sc I}JASS and H{\sc I}PASS for galaxies in the 
 region 12$^{h}$08$^{m}$$\rightarrow$12$^{h}$34$^{m}$, 
  Decl.=22\degr$\rightarrow$25\degr. }
\end{figure}

In recent years several authors have attempted to determine the H{\sc I}MF of the 
local Universe using an H{\sc I}-selected sample of galaxies, with conflicting results.
 For example, using the data from the 
 the Arecibo H{\sc I} Strip  Survey (Sorar 1994),  Zwaan et al. (1997) derived an H{\sc I}MF 
 with a shallow faint end slope ($\alpha$=1.2) consistent with earlier H{\sc I}MFs derived from 
 optically selected samples. In contrast the H{\sc I}MF derived from the  
  Arecibo Slice survey (Schneider, Spitzak \& Rosenberg 1998; Spitzak \& Schnedier 1998)
   has an up-turn in its lowest mass bin, although this is due to only 2 galaxies in this bin.  
 Recently, Rosenberg \& Schneider (2002) have also reported a steep faint end slope 
 ($\alpha$$\simeq$1.5) to the H{\sc I}MF they have derived from the Arecibo Dual-Beam 
 Survey (Rosenberg \& Schneider 2000). 
  H{\sc I}PASS and H{\sc I}JASS will provide much larger samples of galaxies and greatly improve the statistics 
 of such determinations of the H{\sc I}MF.  However, previous studies based upon H{\sc I}-selected samples of galaxies have 
 tended to overlook the important effect that the inclination of a galaxy to the line of sight could have on its inclusion in such a 
 sample. There are two factors to be considered.

Firstly, highly inclined galaxies may suffer from significant self-absorption.
Studies of the H{\sc I} emission from galaxies have generally assumed that the H{\sc I} line is optically thin in all circumstances. 
  Relatively few authors (e.g. Epstein 1964a,b; Haynes \& Giovanelli 1984) 
 have addressed the issue of whether the H{\sc I} emission from galaxies is actually
  optically thin in all galaxies. 
   If this assumption is not valid for highly inclined galaxies, then the H{\sc I} masses
 of such galaxies will have been underestimated. Some highly inclined galaxies will be missed altogether from an H{\sc I}-selected 
 sample of galaxies, despite less inclined galaxies of the same H{\sc I} mass being included.  Both of these 
 effects will lead to errors in the derived H{\sc I}MF.

Secondly, as noted in Section~4.2, there is a minimum believable velocity-width, $\Delta$V$_{\rm 20}^{\rm lim}$,  
 which an object in a blind H{\sc I} survey must have  
  in order to be distinguishable from narrow-band radio frequency interference.
 For any given type of galaxy, the measured velocity-width will be narrower the more face-on the 
 galaxy is. Hence, some galaxies with  inclinations close to the line of sight could be missed.

In this section, we use the sample of confirmed H{\sc I}JASS sources 
 to study the relative seriousness of these two selection effects on the 
 composition of an H{\sc I}-selected sample of galaxies and the implications this has for derivations of the H{\sc I}MF and 
 $\Omega_{\rm HI}$.

\subsection{The expected distribution of galaxies in an H{\sc I}-selected sample as a function of inclination angle}

If the assumption that the 21-cm line of H{\sc I} is always optically thin is correct, then 
  the relationship between integrated H{\sc I} flux from a galaxy, S$_{\rm Int}$ (in Jy~km\,s$^{-1}$),  and total H{\sc I} mass, 
 M$_{\rm HI}$ (in M$_{\odot}$), is  given by  
\begin{equation}
M_{\rm HI}=2.356 \times 10^{5} S_{\rm Int} D^{2} 
\end{equation}
where   D is the distance in Mpc
 (see e.g. Rohlfs 1986). 

Now, if we assume that H{\sc I} emission from any given type of galaxy 
 is not necessarily optically thin, i.e. that  there is an inclination-dependent opacity effect,  
then we can re-write equation (4)  as 
\begin{equation}
M_{\rm HI}= 2.356 \times 10^{5} S_{\rm Int} D^{2} f(i)  
\end{equation}
where f(i) is the correction factor needed to correct the H{\sc I} mass derived from the optically
 thin assumption to the actual H{\sc I} mass. Assuming this function is significant at all, then 
 f(i) may vary for different morphological types and will increase as inclination 
 angle increases.

From eqn (5), it follows that  the integrated flux, S$_{\rm Int}$,  which a galaxy of 
 H{\sc I} mass, M$_{\rm HI}$,  
  and inclination angle, i, will have at 
  a distance of D can  be written as:
\begin{equation}
S_{\rm Int}=  \frac{1}{2.356 \times 10^{5}} . \frac{M_{\rm HI}}{D^{2}}  . \frac{1}{f(i)} 
\end{equation}

Hence, if a survey has an integrated flux limit, S$_{\rm Int}^{\rm lim}$, then the maximum distance, D$_{\rm max}$ in Mpc, 
  at  
 which one could detect a galaxy of H{\sc I} mass, M$_{\rm HI}$, and inclination angle, i, would be given by 
\begin{equation}
D_{\rm max} = \left [  \frac{1}{2.356 \times 10^{5}} . \frac{M_{\rm HI}}{f(i)}  . \frac{1}{S_{\rm Int}^{\rm lim}} 
 \right ]^{\frac{1}{2}}  
\end{equation}

However, as discussed in Section 4.2, 
 the H{\sc I}JASS sample does not have a single S$_{\rm Int}^{\rm lim}$ value. The sample is approximately  peak 
 flux limited and the integrated flux limit varies with velocity-width such that:
\begin{equation}
S_{\rm Int}^{\rm lim} \simeq 0.6 \Delta V_{20} S_{\rm pk}^{\rm lim}
\end{equation}

So, for a given peak flux limit,  the maximum distance at which a galaxy could lie and still be included in the HIJASS sample is related to its M$_{\rm HI}$, $\Delta$V$_{\rm 20}$ and i, by:
\begin{equation}
D_{\rm max} \propto \left [ \frac{M_{\rm HI}}{f(i)}  . \frac{1}{\Delta V_{\rm 20}}  \right ]^{\frac{1}{2}} 
\end{equation}
and the volume, V(i) (in Mpc$^{3}$), within which such a galaxy could lie and still be included within H{\sc I}JASS is 
 related to M$_{\rm HI}$,  $\Delta$V$_{\rm 20}$ and i by:
\begin{equation}
V(i) \propto \left [ \frac{M_{\rm HI}}{f(i)}  . \frac{1}{\Delta V_{\rm 20}}  \right ]^{\frac{3}{2}} 
\end{equation}

Now,  we expect galaxies to be randomly oriented in space. If so then the  intrinsic 
 distribution of galaxies as a function of inclination angle, N(i), is described by 
\begin{equation}
N(i) \propto {\rm sin}\, i
\end{equation}
To find the expected observed distribution of galaxies  as a function of 
 inclination angle, $\phi$(i), we have to multiply the intrinsic distribution  of galaxies as a 
 function of i, N(i), by the volume within which a galaxy at a given i can be observed, V(i), i.e. 
\begin{equation}
\phi(i) \propto  \left [ \frac{M_{\rm HI}}{f(i)}  . \frac{1}{\Delta V_{\rm 20}}  \right ]^{\frac{3}{2}}  . \, sin\,i 
\end{equation}

The expected observed distribution of the HIJASS sample as a function of inclination angle therefore depends on several other 
relationships: the distribution of galaxies as a function of M$_{\rm HI}$; the relationship (if any) between 
 M$_{\rm HI}$ and $\Delta$V$_{o}$; the relationship between $\Delta$V$_{o}$ and $\Delta$V$_{20}$ and 
 inclination angle, i. In Section 5.3 we show that at large i (i.e. $>$50\degr) this relationship can be 
 simplified and used to study the effect of HI self-absorption within the HIJASS sample. Firstly, in Section 5.2, 
 we consider the effect of the velocity-width limit on the HIJASS sample at small inclination angles.

\subsection{Effect of the velocity-width limit on the HIJASS sample}

For a given galaxy, $\Delta$V$_{\rm 20}$ is an observed property which depends on a combination of its 
 rotational velocity, V$_{\rm rot}$, its inclination to the line of sight, i, its internal velocity dispersion, 
 $\Delta$V$_{\rm t}$ (i.e. that due to turbulence and non-planar motions within the galaxy) and the 
 contribution of instrumental broadening to the velocity width, $\Delta$V$_{\rm inst}$. For low M$_{\rm HI}$ galaxies, 
 Tully \& Fouque (1985) showed that these properties can be related via the equation:
\begin{equation}
\Delta V_{\rm 20} = \left[ \Delta V_{o}^2 \, ({\rm sin}\, i)^2  + \Delta V_{t}^2 \right]^{1/2} + \Delta V_{inst}
\end{equation}
 where  $\Delta$V$_{\rm o}$=2V$_{\rm rot}$ is the linewidth the 
 galaxy would have if edge-on (i.e.  ignoring the internal velocity-dispersion).  
 For higher M$_{\rm HI}$ galaxies, the $\Delta$V$_{\rm t}$ term adds linearly to the 
  measured velocity-width (see also Verheijen \& Sancisi 2001) since these galaxies generally show 
  `boxy' HI profiles rather than the typical Gaussian profiles of dwarf galaxies. 
   However, since the velocity-width limit is more important for dwarf than giant galaxies, 
  we conservatively adopt the quadratic summation of eqn(13). 

If we assume that a galaxy cannot be seen by the survey if 
 $\Delta$V$_{\rm 20}$$<$$\Delta$V$_{\rm 20}^{\rm lim}$ then we can express the minimum 
 inclination angle that a galaxy must have in order to be included in the sample as 
\begin{equation}
i_{\rm min} = {\rm arcsin}\left[ \frac{(\Delta V_{20}^{lim} - \Delta V_{inst})^{2} - \Delta V_{t}^{2}}{\Delta V_{o}^{2}} \right]^{1/2}
\end{equation}

 To illustrate the effect of the velocity-width limit, we adopt the 
 value $\Delta$V$_{\rm t}$=20$\pm$2~km\,$^{-1}$ found by Rhee (1996) from a study of 
 28 galaxies with well defined HI velocity fields. This value is in close agreement with 
 those found by similar studies by Broeils (1992) and Verheijen \& Sancisi (2001).  We use the 
 Bottinelli et al. (1990) estimate of $\Delta$V$_{\rm inst}$ = 0.55 $\times$ {\it R}, where 
 {\it R} is the velocity resolution of the survey. This gives 
 $\Delta$V$_{\rm inst}$=10~km\,s$^{-1}$ for HIJASS.
 We assume $\Delta$V$_{20}^{lim}$=52.8~km\,s$^{-1}$ (see Section 4.2).

Table~4 illustrates the effect of the velocity-width cut-off on the number of galaxies included in 
the HIJASS sample for a range of $\Delta$V$_{o}$ values (Column 1). Column 2 lists the 
 i$_{min}$ for each $\Delta$V$_{o}$, found using eqn(14) and our assumed values above. 
 As expected this effect gets progressively more serious for inherent narrow velocity-width 
 objects: galaxies with $\Delta$V$_{o}$$<$100~km\,s$^{-1}$ cannot be seen with i$<$22\degr; galaxies with 
 $\Delta$V$_{o}$$<$50~km\,s$^{-1}$ are missed if i$<$49\degr.

\begin{table}
\caption{Illustration of the effect of the velocity-limit selection effect 
 on the inclusion of galaxies within the HIJASS sample. 
  Column 1 is a set of $\Delta$V$_{o}$ values. Column 2 lists the minimum 
 inclination angle, i$_{\rm min}$, which a galaxy of each $\Delta$V$_{o}$ 
 must have in order to be included in the sample (using eqn 14). Column 3 
 lists $\zeta$, the fraction of galaxies which will be missed from the sample at each 
 $\Delta$V$_{o}$.}
\begin{tabular}{lrr} \hline
$\Delta$V$_{o}$  & i$_{min}$  & $\zeta$ \\
km\,s$^{-1}$ & deg & per cent \\  \hline
40 &  63.5\degr   &  55.4  \\
50 &  49.2\degr & 34.6 \\
75 &  30.3\degr & 13.7 \\
100 & 22.2\degr &  7.4 \\
150 & 14.6\degr &  3.2\\
200 & 10.9\degr & 1.8\\
250 &  8.7\degr & 1.2\\
300 &  7.3\degr & 0.8\\
350 &  6.2\degr & 0.6 \\ 
400 &  5.4\degr & 0.5\\ \hline
\end{tabular}
\end{table}

The actual fraction of galaxies missed at each $\Delta$V$_{o}$ can be found by integrating from 
 i=0 to i=i$_{min}$ over a randomly oriented sample (see Zwaan et al., in preparation):
\begin{equation}
\zeta = \int_{o}^{i_{min}} sin\,i \,\, di  = 1 - cos\,i_{min}
\end{equation}
The derived values of $\zeta$ are listed in Column 3 of Table~4. Clearly, within HIJASS data, this selection effect
 becomes progressively more important at smaller $\Delta$V$_{o}$. Whilst only 7\% of galaxies 
 have been missed at $\Delta$V$_{o}$=100~km\,s$^{-1}$, this number has risen to 35\% at 
 $\Delta$V$_{o}$=50~km\,s$^{-1}$. 

 As noted above, we cannot properly consider the effect that the velocity-width cut-off may have on a derived 
 HIMF without knowing whether there is a relationship between $\Delta$V$_{o}$ and M$_{\rm HI}$ and, if 
 so, what form that relationship takes. 
 Whilst one could argue about the precise relationship between M$_{\rm HI}$ and $\Delta$V$_{o}$, 
  previous studies suggest that the two are related such that more massive galaxies appear to have broader 
 velocity-widths. For example, 
 from H{\sc I} measurements of an optically-selected sample of galaxies, Rao \& Briggs (1993) derived
 $\Delta$V$_{\rm 20}$=0.15M$_{\rm HI}^{1/3}$.
    From their H{\sc I}DEEP sample, Minchin et al.(in preparation) have found that 
 $\Delta$V$_{\rm 20}$=0.42M$_{\rm HI}^{0.28}$. However, such relations may be partly 
 due to selection effects (see e.g. Minchin 2001). 

Fig.~12 is a  plot of $\Delta$V$_{o}$ against M$_{\rm HI}$ for 186 H{\sc I}JASS galaxies for which we have derived inclination angles.
 This sample includes all of the previously catalogued galaxies
 (expect those listed as `pair' or `group'), the 15 ASSs  for which the 
 optical identification was relatively unambiguous (i.e. not the ASS*s) and the 5 PUGs 
 for which an obvious optical counterpart could be seen on the DSS. The inclination for each galaxy was determined from 
 the ratio of the semi-major to the semi-minor axis using the  equation:
\begin{equation}
{\rm cos}^2 i = \frac{(b/a)^2 - r_{o}^{2}}{1 - r^{2}_{o}}
\end{equation}
(Holmberg 1958), where r$_{o}$ is the intrinsic axis ratio of an edge-on disk. Estimates for 
 r$_{o}$ vary between 0.11 and 0.2 for this property. We have assumed a value of 0.16 for every galaxy. 
  Note that this may be significantly inaccurate 
 for low-mass dwarf galaxies, for which Staveley-Smith, Davies \& Kinmann (1992)  found that 
 values up to about 0.5 may be appropriate. However, relatively few of the galaxies in 
 the HIJASS sample are dwarfs (see Fig.~8a). 
 The values of b/a were taken from the Third Reference Catalogue of Bright Galaxies 
  (de Vaucouleurs et al. 1991) where available. Otherwise they were determined 
 from DSS images of each galaxy using the SExtractor package (Bertin \& Arnouts 1996).

Included on Fig.~12 is the locus of a line showing the relationship
\begin{equation}
\Delta V_{\rm o} = 0.42\, M_{\rm HI}^{0.3}
\end{equation}
which gives the best fit to our data. Note, however, that there is a wide scatter about this locus. There are
   very few data points at 
 M$_{\rm HI}$$<$10$^{8}$~M$_{\odot}$ and many of these lie a long way from the locus of eqn 17. 
  Hence, the conclusions we draw using this relationship, especially at low M$_{\rm HI}$,  
    should be treated as only illustrative of the possible effect 
 of ignoring the velocity-width limit selection effect.

\begin{figure}
  \epsfig{file=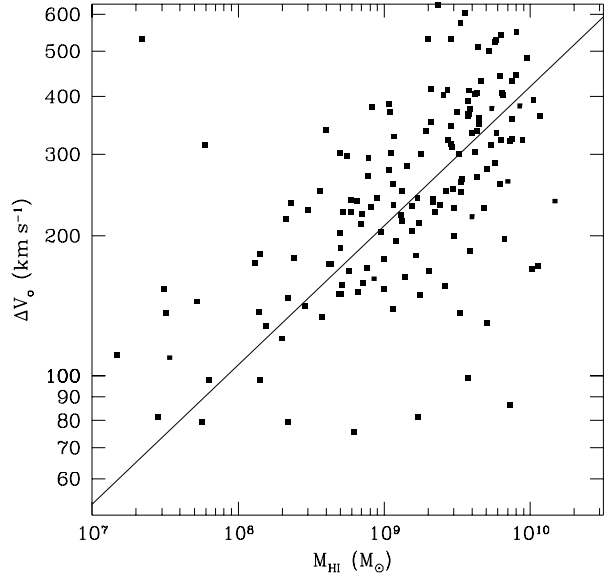}
  \caption{Plot of log($\Delta$V$_{o}$) against log(M$_{\rm HI}$) for those 186 H{\sc I}JASS galaxies
 for which we have inclination angles. The solid line is the locus of the relationship 
 $\Delta$V$_{o}$=0.42M$_{\rm HI}^{0.3}$ (equation 20). } 
\end{figure}

 Table~5  lists the $\Delta$V$_{\rm o}$ (Column 2) equivalent to a range of M$_{\rm HI}$ values (Column 1), assuming the 
 $\Delta$V$_{o}$-M$_{\rm HI}$ relationship of eqn(17). 
  Also listed (Column 3) is 
  the  minimum inclination angle i$_{\rm min}$ which a galaxy of each $\Delta$V$_{o}$ could have and still be included 
 in the H{\sc I}JASS sample (assuming $\Delta$V$_{\rm 20}^{lim}$=52.8~km\,s$^{-1}$) (from eqn.14). 
 Column 4  lists the fractional error, $\zeta$,   
  which would be introduced into the H{\sc I}MF at each mass as a result of the exclusion from the  
  H{\sc I}-selected sample of galaxies at i$<$i$_{min}$ (from eqn.15).
.

\begin{table}
\caption{Illustration of the effect of the velocity-limit selection effect on the H{\sc I}MF 
 derived  from  H{\sc I}JASS data
  ($\Delta$V$_{\rm 20}^{\rm lim}$=52.8~km\,s$^{-1}$). Column 2 shows the $\Delta$V$_{\rm o}$ 
 value for a range of M$_{\rm HI}$ values (Column 1), derived assuming eqn (17) (note the 
 caveats about this in the main body of the text). 
  Column 3 shows the minimum inclination angle, i$_{min}$, at which a galaxy of this 
 $\Delta$V$_{o}$ would be included in the sample (from eqn 14). Column 4 shows the percentage error 
  in the H{\sc I}MF at each mass resulting from excluding galaxies with i$<$i$_{min}$ (from 
 eqn~15)  from the sample. }
\begin{tabular}{lrrr} \hline
M$_{\rm HI}$ & $\Delta$V$_{o}$ & i$_{min}$ & $\zeta$ \\ 
 M$_{\odot}$ & km\,s$^{-1}$ & deg &  per cent\\  \hline
2$\times$10$^{7}$ &   65  & 35.6  & 18.7   \\
5$\times$10$^{7}$ &   86  & 26.1  & 10.2 \\
1$\times$10$^{8}$ &   105 & 21.1  & 6.7 \\
5$\times$10$^{8}$ &   171 & 12.8   & 2.5 \\ 
1$\times$10$^{9}$ &   210 & 10.4  & 1.6 \\
5$\times$10$^{9}$ &   341 &  6.4 & 0.6 \\
1$\times$10$^{10}$ &  420 &  5.2  & 0.4 \\ \hline
\end{tabular}
\end{table}
 
 As is clear from Column 4 of Table 5, 
  under these assumptions, for the H{\sc I}JASS sample, we would be significantly 
 underestimating the H{\sc I}MF at 
 M$_{\rm HI}<$10$^{8}$~M$_{\odot}$ if we did not compensate for the velocity-width limit 
 selection effect. What is most striking is that the extent of our 
 underestimate of the H{\sc I}MF would increase  
  at the low M$_{\rm HI}$/small $\Delta$V$_{o}$ end. This implies that H{\sc I}MFs 
 derived from H{\sc I}-selected samples without correcting for this effect may have significantly
  underestimated the steepness of the  faint-end slope. 
  There are, however, other complicating factors which may affect the low mass end, e.g.
    the relationship of $V_{\rm rot}$ to velocity dispersion in dwarfs
   (e.g. Lo, Sargent \& Young 1993, Staveley-Smith et al. 1992). At best, Table~5 is a warning that 
 a consideration of the effect of the velocity-width cut-off on sample completeness should be an 
 essential part of any derivation of the HIMF.

\subsection{Effect of HI self-absoprtion}

To study the possible impact of HI self-absorption on the HIJASS sample, we ideally wish to study the 
 actual observed distribution of inclination angles of the sample against that predicted for a 
 sample with no HI self-absorption. Eqn (12) describes the expected observed distribution 
 of galaxies as a function of inclination angle, $\phi$(i). As noted above, this is a complex function 
 which depends on the relationship betwwen $\Delta$V$_{o}$ and M$_{\rm HI}$ and that between 
 $\Delta$V$_{o}$ and $\Delta$V$_{20}$. However, if we restrict our analysis to large inclination 
 angles then we can reasonably make two simplifying assumptions. The first is that we are not 
 missing a significant number of galaxies due to the velocity-width limit 
  selection effect. As noted in Table~4, only 
 a galaxy with $\Delta$V$_{o}$$<$50~km\,s$^{-1}$ will be missed due to this effect at i$<$50\degr. 
 The second is that at i$>$50\degr the thermal velocity dispersion, V$_{t}$, 
  no longer makes a significant contribution to V$_{20}$ and, hence, we can assume that 
\begin{equation}
\Delta V_{\rm 20} \simeq \Delta V_{\rm o} \, sin\,i
\end{equation}

In this case, the expected observed distribution of galaxies as a function of i can be written as
\begin{equation}
\phi(i) \propto    \left [ \frac{M_{\rm HI}}{\Delta V_{\rm o}} \right ]^\frac{3}{2} . \left [ \frac{1}{f(i)}  \right ]^{\frac{3}{2}}  .  \left [ \frac{1}{sin\,i} \right ]^\frac{1}{2} 
\end{equation}

Since M$_{\rm HI}$ and $\Delta$V$_{o}$ are constant for a given galaxy, the expected
  observed distribution of all galaxies in 
 the H{\sc I}JASS sample as a function 
 of inclination  angle can then be described by 
\begin{equation}
\phi(i) \propto  f(i)^{-1.5}  ({\rm sin}\,i)^{-0.5}   
\end{equation}
 The implication 
 of this equation is that, in the absence of significant self-absorption, 
  we expect to see a relatively flat distribution at i$>$50$^{o}$.

Fig.~13 presents a histogram of the derived inclination angles for those of the 186 HIJASS galaxies
 for which we have derived inclination angles where i$>$50\degr (105 galaxies). 
  In the optically thin scenario, we expect a very shallow fall-off in the observed distribution 
   at high inclination angles. 
 We actually observe  a sharp fall in the observed number of galaxies at i$>$74\degr. This
 could be  because self-absorption 
 becomes significant in at least some galaxies at these high inclinations.

\begin{figure}
  \epsfig{file=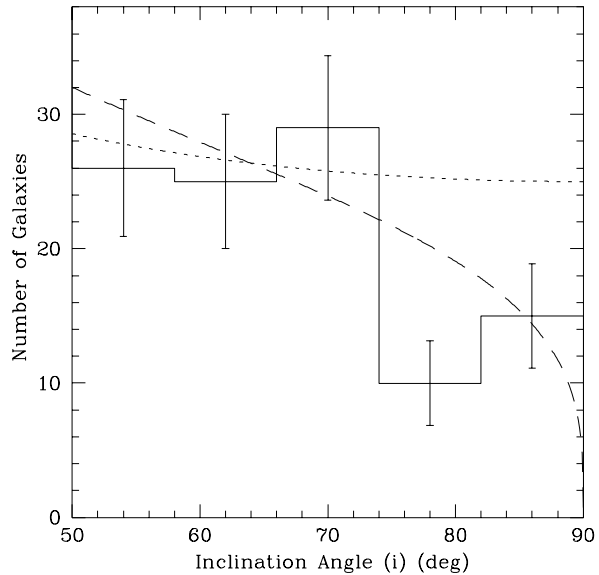}
  \caption{Histogram showing observed distribution of H{\sc I}JASS galaxies as a function of inclination angle, i, 
 for galaxies with i$>$50\degr. Each bin has width 8\degr. 
   The short-dashed line shows a $\phi$(i) function, fitted to the observed distribution 
 in the range i=50\degr$\rightarrow$74\degr\, (assumes no self-absorption).  
 The long-dashed line shows the best fitting $\phi$(i) function to galaxies in the range 
  i$>$50\degr\, (assumes self-absorption with a best fit value of $\beta$=0.2).}
\end{figure}

 We can quantify 
 the effect of self-absorption if we assume that the sample of galaxies in the 
 range i=50\degr$\rightarrow$74\degr is free from the effects of 
 both the velocity-width cut-off  and HI self-absorption.  Note that the velocity-width limit will tend 
 to flatten the observed distribution as 
 a function of i, so this assumption may lead to us underestimating the effects of self-absorption rather 
 than over-estimating it. 
 We have fitted a  $\phi$(i) function to the  observed distribution in the 
 range  i=50\degr$\rightarrow$74\degr. The best fit 
  was determined  by normalising $\phi$(i) such that the theoretical number of galaxies
  in the range i=50\degr$\rightarrow$74\degr\, is equal to the observed number of galaxies in this range. 
  The best fitting $\phi(i)$ function  is plotted 
  on Fig.~13 (short-dashed line).

 This best fitting $\phi$(i) distribution predicts 
 that there should be 51$\pm$7 galaxies at i$>$74\degr. This compares to the observed number of 
 25 galaxies, i.e. a 3.5$\sigma$ shortfall of galaxies. An alternative fit can be made by normalising
  the  theoretical $\phi$(i) 
 function in the range i=50\degr$\rightarrow$74\degr\, to 1$\sigma$ below the total observed counts 
 in this range. Such a best fit predicts that there should be a total of 44$\pm$7 galaxies 
 at i$>$74\degr, still 2.5$\sigma$ above the number observed.

The largest previous study of this issue was that of Haynes \& Giovanelli (1984) who obtained H{\sc I} 
 measurements 
 of 288 isolated galaxies using the Arecibo 305-m telescope. They compared the HI surface density (defined 
as the ratio of integrated H{\sc I} flux  to  optical surface area of the galaxy)  with the axial ratio 
 for the galaxies as a function of morphological type. They found that for Sa, Sab, Sb, Sbc and Sc galaxies 
 there was a clear trend for the measured surface density to fall as inclination to the line of sight 
 increases.  
 The implication of this is that the measured column depth of a highly inclined galaxy is less than it 
would be 
  for more face-on objects because a fraction of the H{\sc I} is being self-absorbed.
 Haynes \& Giovanelli found a general tend for f(i) [the inclination-dependent H{\sc I} mass correction 
 factor - see eqn (5)]
to vary as 
\begin{equation}
f(i)=(cos i)^{-\beta}
\end{equation} 
 where  $\beta$ is a constant dependent on morphological type. They found values of 
  $\beta$ of 0.04 for Sa and Sab, 0.16 for Sb, and 0.14 for Sbc and Sc galaxies. 
 They found no corrections to be necessary for galaxies earlier than Sa or later 
 than Sc, indicating self-absorption to be negligible in these types.

We adopt a similar form for f(i) and used a $\chi^{2}$ minimisation technique
   to  derive the value of 
 $\beta$ which gives a best fit to our observed distribution at i$>$50\degr. 
 This best fitting value is $\beta$=0.2. This ignores the possible dependence of 
 $\beta$ on morphological type. 
 The $\phi$(i) function derived 
 using $\beta$=0.2 is shown on Fig.~13 (long-dashed line). This value is significantly 
 larger than the largest value derived by Haynes \& Giovanelli (1984). 
 Note, however, that our model does not provide a particularly good fit to the data above i=74\degr, 
 especially in the bin centered at i=78\degr. This may be a consequence of the relatively small 
 total number of galaxies in our sample or of the assumed functional form of f(i) not being 
 appropriate. 

 Using the argument of Zwaan et al. (1997), the average  
 effect of self-absorption on measured M$_{\rm HI}$ can be obtained by averaging f(i) over a 
random distribution of inclinations. 
\begin{equation} 
<f(i)> = \frac{\int_{o}^{\pi/2} (cos\,i)^{-\beta} sin\,i \,di}{\int_{o}^{\pi/2} sin\,i\, di}    =  \frac{1}{1 - \beta}
\end{equation}
giving a mean correction over all inclinations of $<$f(i)$>$=1.25 for $\beta$=0.2. We have no knowledge of how f(i) 
 varies with M$_{\rm HI}$ or morphological type. If it is uncorrelated with M$_{\rm HI}$ then the effect of this 
 on the H{\sc I}MF would be to shift galaxies of each mass to  higher masses by an average factor of 1.25. This would lead to 
 a corresponding increase in M$_{\rm HI}^{\star}$. The shape of the H{\sc I}MF would be unaltered.

 This correction factor of 1.25 to M$_{\rm HI}^{\star}$  is a lower limit for two reasons. 
 Firstly, we found a best fitting   $\beta$-corrected $\phi$(i) function by assuming
   that the velocity-width limit effect was not significant at 
 i$>$50\degr. As is clear from Table~4, some intrinsically narrow velocity-width  galaxies will be lost 
 even at i$>$50\degr. Hence, our value of $\beta$ is a lower limit. 
  Secondly, in deriving $<$f(i)$>$ we have averaged over all i.
 We should actually integrate over i=i$_{min}\rightarrow$90\degr\, for any given $\Delta$V$_{o}$ (since galaxies 
 at i$<$i$_{min}$ will not have been included in the sample). This would have the effect of 
 increasing $<$f(i)$>$ for those galaxies actually included in the sample. 

We also have to account for the fact that self-absorption is not only causing us to underestimate the 
 mass of some galaxies, but is also  causing some galaxies to be excluded 
   from the sample altogether. In a randomly oriented sample of galaxies, 
   28 per cent of the contribution to the H{\sc I}MF should come 
 from galaxies with i$>$74\degr. However, at i$>$74\degr\, we are missing at least 40 per cent of those galaxies 
   which we would expect to see in the absence of self-absorption. If we assume that 
 self-absorption is not correlated with $\Delta$V$_{o}$ or M$_{\rm HI}$ then this effect will 
 cause us to underestimate the H{\sc I}MF by a factor of 1.14 at each M$_{\rm HI}$, i.e. to derive the correct 
 H{\sc I}MF we would need to correct the normalisation $\theta$* by a factor of at least 1.14.  
 The combined effect of the correction factors of 1.25 in M$_{\rm HI}^{\star}$ and 1.14 in $\theta$* would be to increase the 
 derived value of $\Omega_{\rm HI}$ by at least 25 per cent.

Experience in the optical suggests that correcting for the number of self-absorbed discs in a 
 survey, using only those you can see, is extremely model-dependent (Disney, Davies \& Phillipps 
 1989; Witt, Thronson \& Capuano 1992). For instance, in the present
  case the optical depth could vary by an order
 of magnitude  as between flat and solid rotation curves. All we can truly say for now 
  is that an HI-select survey like ours would, of all surveys, be most likely to run 
  into HI self-absorption, and that the affect is 
 plainly significant. How significant remains a question for the future.

\section{Concluding Remarks}

 This paper has described the properties of the  present sample of confirmed sources derived from the H{\sc I}JASS data.  
 This sample  will be added to as further 
 sources are confirmed. This will obviously follow further observing runs on the Lovell 
 telescope. However, 
 there is  scope within existing H{\sc I}JASS data for adding to the present sample. 
 As noted in Section~3.1, the `possible' detections from the 2002 run have not yet 
 been followed up by single-beam narrow-band observations. Those confirmed in this 
 way will then be added to the sample.
 We noted in Section 4.2 the important selection effect that, because our sample is 
 effectively peak flux limited, the integrated flux limit 
   is a function of $\Delta$V$_{\rm 20}$. This leads to a major bias against 
 galaxies with broad velocity-widths (and presumably higher H{\sc I} masses). We are developing alternative 
 detection techniques with the aim of detecting broader velocity-width sources to similar integrated 
 flux levels as we can presently detect narrow-line sources.
 There is a further possibility that nearby spatially
 extended objects may be removed from the data by the conventional bandpass correction 
 algorithm used by {\sc LIVEDATA}. Alternative algorithms  are being tested to determine if any sources
 have been lost in this way. We are also considering ways of detecting  
  massive galaxies in the data between $cz$=7500--9000~km\,s$^{-1}$. 
 It is possible that any galaxies massive enough to be detectable at this distance
   will 
 have very broad velocity-widths and hence, be hard to actually detect despite their mass.

We have embarked upon a detailed follow-up program in order to 
  fully study the astrophysical nature of the  objects within the H{\sc I}JASS
 sample: in particular the previously uncatalogued objects. 
As discussed in  Section~1, 
 the addition  of these previously uncatalogued objects  to the extragalactic 
 census of the local Universe  
  could have a significant impact not only on determinations of the luminosity density 
 and mass density of the local 
 universe but also on our understanding of the processes of galaxy formation and evolution. 
 They could be systems which have undergone a very different 
 formation process and/or evolutionary path than optically selected 
 galaxies. Are they old galaxies which have evolved slowly and 
 have yet to transmute most of their gas into stars ? Or are they
 young objects which are still at an early stage of their star 
 formation histories ? Are they objects which have recently accreted large 
 amounts of H{\sc I}~? To determine this we require information on their 
 morphological and structural properties; on their stellar populations; and 
 on their star formation rates, star formation 
 histories and metalicities.
All of the potential new detections will be observed using the Westerbork 
 Synthesis Radio Telescope (WSRT) in order to get accurate positions for these 
 objects and to map the distribution of H{\sc I} within them. This will also enable us to 
  decide which of the 23 objects lying close to 
 a catalogued galaxy with no measured redshift (ASSs) are  actually 
 associated with that  previously 
catalogued galaxy and which are new objects. 
 Broad band imaging 
  is  being undertaken with the Isaac Newton Telescope of the most 
 interesting objects. 
 We are attempting to detect H$\alpha$ emission from the nearer of the objects 
 using the Jakobus Kapteyn Telescope. We have also been awarded time 
 on the William Herschel Telescope to obtain IR imaging of several of the objects. 
Future publications will present the results of  this follow-up program.

We have shown, using the H{\sc I}JASS sample, that self-absorption is a significant, but often overlooked, effect in galaxies at high inclinations.
Properly accounting for it could increase the derived H{\sc I} mass density by at least 25 per cent
 and possibly a lot more. The 
  effect this will have on the shape 
 of the H{\sc I}MF will depend on how self-absorption affects galaxies of different $\Delta$V$_{o}$.  

We have also shown that the 
 velocity-width limit will always act so as to exclude low inclination angle galaxies from H{\sc I}-selected samples. This 
  affect will become progressively more serious at lower $\Delta$V$_{o}$ values. If, as we might expect, galaxies with smaller
 intrinsic velocity-widths have smaller H{\sc I} masses, then compensating for this effect 
 could significantly steepen the faint 
 end of the H{\sc I}MF.

\section*{Acknowledgments}

The construction of the multibeam receiving system at Jodrell 
Bank was made possible by the UK PPARC (Grant No. GR/K28237 to Jodrell Bank). 
  The HI Jodrell All Sky Survey is  funded by the UK 
PPARC (Grant No. PPA/G/S/1998/00620 to Cardiff). Both grants are gratefully acknowledged. 
PJB and RFM also acknowledge the financial support of the UK PPARC. 
We thank the Director of the Jodrell Bank Observatory, Prof. 
 Andrew Lyne,  for granting 
 the observing time for H{\sc I}JASS and for making available the facilities of the 
 observatory. 
 We thank CSIRO Radiophysics in Australia for producing much of the design and 
 engineering of the system.  
We also thank the following for their help with H{\sc I}JASS: Gareth Banks, 
 Dave Brown, Mark Calabretta, Jim Cohen, Jon Davies, 
  Rod Davies, Judy Haynes, Anthony Holloway, Ian Morison, Rhys Morris, 
  Rodney Smith, Lister Staveley-Smith, Daniel Zambonini  and the staff and students 
 of the Jodrell Bank  Observatory. 
 This research has made use of the NASA/IPAC Extragalactic Database (NED) which is
operated by the Jet Propulsion Laboratory, Caltech, under agreement with the 
 National Aeronautics and Space Administration. This research has also made use of the Digitised 
 Sky Survey, produced at the Space Telescope Science Institute under US Government Grant NAG W-2166.


\begin{thebibliography}{99}

\bibitem{b} Banks G.D., et al., 1999, ApJ, 524, 612
\bibitem{b} Barnes D.G. et al., 2001, MNRAS, 322, 486
\bibitem{b} Baugh C., Cole S., Frenk C.S., 1996, MNRAS, 283, 1361
\bibitem{b} Bertin E., Arnouts S., 1996, A\&AS, 117, 393
\bibitem{b} Bell E.F., Bower R.G., 2000, MNRAS, 319, 235
\bibitem{b} Bell E.F., de Jong R.S., 2000, MNRAS, 312, 497 
\bibitem{b} Bird T.S., 1994, in IEEE Antennas \& Propagation Symposium.  Seattle, p.966
\bibitem{b} Bird T.S., 1997, in IEEE Antenna \& Propagation Society Symposium, Montreal, p. 1618
\bibitem{b} Bottinelli L., Gouquenheim L., Fouque P., Paturel G., 1990, A\&AS, 82, 391
\bibitem{b} Boyce P.J. et al., 2001, ApJ, 560, L127 
\bibitem{b} Broeils, A.H., 1992, PhD Thesis, Univ Groningen
\bibitem{b} Canaris J., 1993, in Workshop on New Generation Digital Correlators. Publisher, Tuscon, p.117 
\bibitem{b} Cross N.J.G., et al., 2001, MNRAS, 324, 825
\bibitem{b} Disney M.J., 1976, Nature, 263, 573
\bibitem{b} Disney M.J., 1999, in Davies J.I., Impey C., Phillipps S., eds, Low Surface Brightness Universe, ASP, San Francisco
\bibitem{b} Disney M.J., Phillipps S., 1987, Nature, 329, 203
\bibitem{b} Disney M.J., Davies J.I., Phillipps S., 1989, MNRAS, 239, 939 
\bibitem{b} Epstein E.E., 1964a, AJ, 69, 490
\bibitem{b} Epstein E.E., 1964a, AJ, 69, 512
\bibitem{b} Freedman W.L. et al., 1994, ApJ, 427, 628
\bibitem{b} Gooch, R,  1995, in ASP Conf. Ser. 77: ADASS IV, Volume 4
\bibitem{b} Haynes M.P., Giovanelli R., 1984, 89, 758
\bibitem{b} Holmberg E., 1958, Medd.Lunds.Astr.Obs., Ser.II, No.136 
\bibitem{b} Impey C.D., Bothun G.D., 1997, ARA\&A, 35, 267   
\bibitem{b} Impey C.D., Bothun G.D, Malin D.F., 1988, ApJ, 330, 634
\bibitem{b} Kauffmann G., Nusser A., Steinmetz M., 1997, MNRAS, 286, 795 
\bibitem{b} Kilborn V.A., 2000, PhD thesis, Univ Melbourne
\bibitem{b} Kilborn V.A. et al., 2000, AJ, 120, 1342
\bibitem{b} Kilborn V.A. et al., 2002, AJ, 124, 690
\bibitem{b} Lo K.Y., Sargent W.L.W., Young K., 1993, AJ, 106, 507
\bibitem{b} McGaugh S.S., 1996, MNRAS, 280, 337
\bibitem{b} Minchin R.F., 2001, PhD thesis, Univ. Wales, Cardiff
\bibitem{b} Phillipps S., Disney M.J., Kibblewhite, E.,  Cawson M.G.M., 1987, MNRAS, 229, 505
\bibitem{b} Rao S., Briggs F.H., 1993, ApJ, 36, 267
\bibitem{b} Rhee, M.-H., 1996, A\&AS, 115, 407
\bibitem{b} Rohlfs K., 1996, Tools of Radio Astronomy, Springer-Verlag
\bibitem{b} Rosenberg J.L., Schneider S.E., 2002, ApJ, 567, 247
\bibitem{b} Rosenberg J.L., Schneider S.E., 2000, ApJS, 130, 177
\bibitem{b} Ryan-Weber E. et al., 2002, AJ, 124, 1954
\bibitem{b} Ryder S.D. et al., 2001, ApJ, 555, 232
\bibitem{b} Sault R.J., Teuben P.J., Wright M.C.H., 1995, ADASS, p.433, ASP  San Francisco
\bibitem{b} Schneider S.E., Spitzak J.G., Rosenberg J.L., 1998, ApJ, 507, L9
\bibitem{b} Solanes J.M., Giovanelli R., Haynes M.P., 1996, ApJ, 461, 609
\bibitem{b} Sorar E., 1994, PhD Thesis, Univ Pittsburgh
\bibitem{b} Spitzak J.G., Schneider S.E., 1998, ApJS, 119, 159
\bibitem{b} Staveley-Smith L., Davies R.D., Kinmann T.D., 1992, 258, 334
\bibitem{b} Staveley-Smith L. et al., 1996, Proc. Astron. Soc. Australia, 13, 243
\bibitem{b} Tully R.B., Fouque., 1985, ApJS, 58, 67
\bibitem{b} de Vaucouleurs G., de Vaucouleurs A., Corwin H.G., Buta R.J., Paturel G., Fouque P., 1991, Third Reference Catalogue of Bright Galaxies, Springer-Verlag, New York
\bibitem{b} Verheijen M.A.W., Sancisi R., 2001, A\&A, 370, 765
\bibitem{b} Witt A., Thronson H.A., Capuano J.M., 1992, ApJ, 393, 611
\bibitem{b} Zwaan M.A., Briggs F.H., Sprayberry D., Soror, E., 1997, ApJ, 490, 173 





\end{thebibliography}
\end{document}